\documentclass[pre,twocolumn,superscriptaddress,showpacs,longbibliography,
nofootinbib]{revtex4-2}

\usepackage{xcolor} 
\usepackage{enumitem}
\usepackage{graphicx}
\usepackage{graphicx,amsmath,amssymb}
\usepackage{bm}
\usepackage[normalem]{ulem}
\usepackage{hyperref}

\usepackage{amsthm}

\usepackage{xr}

\newcommand{\x}{{\mathbf x}}{}
\renewcommand{\u}{{\mathbf u}}
\newcommand{\e}{{\mathbf e}}{}

\newcommand{\tw}{\tilde}{}
\newcommand{\p}{\partial}{}
\renewcommand{\div  }{ \nab \cdot} 
\newcommand{\nab  }{ \nabla} 
\newcommand{\px  }{\p_x}
\newcommand{\pt  }{\p_t}
\newcommand{\dd}{\mathrm{d}}{}
{}

\newcommand{\rhoout}[2]{\bar \varrho_{#1}^{(#2)}}

\newcommand{\pas}{\rho_0}

\newcommand{\pol}{{m}}
\newcommand{\poltravel}{\mathrm{m}}
\newcommand{\ds }{d_s}
\newcommand{\DD }{{\cal D}}

\newcommand{\Dx }{D_T}
\newcommand{\Dt }{D_R}
\newcommand{\vo }{v_0}
\newcommand{\pe}[0]{\text{Pe}}

\newcommand{\h}{\ell_y}

\newcommand{\methodsref}{Methods}

\begin{document}

\title{
Dynamical patterns and nonreciprocal effective interactions in an active-passive mixture through exact hydrodynamic analysis
}%

\author{James Mason}
\affiliation{DAMTP, Centre for Mathematical Sciences, University of Cambridge, Wilberforce Road, Cambridge CB3 0WA, UK}
\author{Robert L. Jack}
\affiliation{DAMTP, Centre for Mathematical Sciences, University of Cambridge, Wilberforce Road, Cambridge CB3 0WA, UK}
\affiliation{Yusuf Hamied Department of Chemistry, University of Cambridge, Lensfield Road, Cambridge CB2 1EW, UK}
\author{Maria Bruna}
\affiliation{DAMTP, Centre for Mathematical Sciences, University of Cambridge, Wilberforce Road, Cambridge CB3 0WA, UK}
\affiliation{Mathematical Institute, University of Oxford, Oxford OX2 6GG, UK}

\begin{abstract}
The formation of dynamical patterns is one of the most striking features of nonequilibrium physical systems. 
Recent work has shown that such patterns arise generically from forces that violate Newton's third law, known as nonreciprocal interactions.  
These nonequilibrium phenomena are challenging for modern theories.
Here, we introduce a model mixture of active (self-propelled) and passive (diffusive) particles amenable to exact mathematical analysis.  
We exploit state-of-the-art methods to derive exact hydrodynamic equations for the particle densities,  which reveal effective nonreciprocal couplings between the active and passive species.  
We study the resulting collective behavior, including the linear stability of homogeneous states and phase coexistence in large systems.  
This reveals a novel phase diagram with the spinodal associated with active phase separation protruding through the associated binodal, heralding the emergence of dynamical steady states.  
We analyze these states in the thermodynamic limit of large system size, showing, for example, that sharp interfaces may travel at finite velocities, but traveling phase-separated states are forbidden.  
The model's mathematical tractability enables precise new conclusions beyond those available by numerical simulation of particle models or field theories.
\end{abstract}

\maketitle

\renewcommand{\thefootnote}{\arabic{footnote}} 

\section*{Introduction}
Simple systems of interacting particles (or agents) can support complex emergent behavior, including the self-assembly of nanoscale equilibrium structures, the self-organization of animals into flocks and swarms, and pattern formation in chemical reactions.
Describing these effects has been a long-standing challenge for physics and mathematics: modern theories focus on emergent nonequilibrium behavior, making it difficult to predict macroscopic collective phenomena from the underlying microscopic rules.
Recent studies have highlighted that nonreciprocal interactions in nonequilibrium systems lead generically to pattern formation in a variety of physical settings~\cite{fruchartNonRecip2021}, including reaction-diffusion systems \cite{braunsBulksurfaceCouplingIdentifies2021,halatekSelforganizationPrinciplesIntracellular2018,rammColiMinCDESystem2019}, living chiral crystals \cite{huangDynamicalCrystallitesActive2020,tanOddDynamicsLiving2022} and quorum sensing bacteria \cite{mukherjeeBacterialQuorumSensing2019,PhysRevLett.131.228302}.  
Despite their diversity, these systems appear to self-organize according to a common set of physical principles, offering the opportunity for a predictive theory with broad scope.

Within this context, the nonreciprocal Cahn--Hilliard (NRCH) equation has recently emerged as a canonical model for nonreciprocally coupled particle models~\cite{youNonreciprocityGenericRoute2020,sahaScalarActiveMixtures2020,frohoff-hulsmannSuppressionCoarseningEmergence2021}.
It illustrates that phase-separated systems can undergo exceptional phase transitions when subjected to nonreciprocal driving, leading to dynamical (traveling) steady states. 
This simple and elegant equation bridges the established equilibrium theory of phase separation and the complex world of pattern formation in nonequilibrium systems. 
However, pattern formation in nonreciprocally coupled systems continues to challenge our understanding, including the vital question of which patterns will appear in any given system~\cite{braunsPhaseSpaceGeometryMassConserving2020,braunsNonreciprocalPatternFormation2023,fruchartNonRecip2021,suchanekIrreversible2023}.

This work addresses these challenges by analyzing a specific nonreciprocal system for which exact mathematical results can be derived. 
Specifically, we introduce an idealized mixture of interacting active and passive particles and derive its exact hydrodynamic limit, the equation that governs its large-scale collective behavior.
Such mixtures are known to display many features of nonreciprocally interacting systems and NRCH-like equations that approximate their motion have been proposed, either on phenomenological grounds or by various approximation schemes \cite{wittkowskiNonequilibriumDynamicsMixtures2017,Dinelli2023,youNonreciprocityGenericRoute2020,cates2024review}. 
These capture many qualitative features of the resulting dynamics but are only partially quantitative.
Our hydrodynamic equation leads to phenomenology similar to that of NRCH and is consistent with generic principles of self-organization via nonreciprocity.

Our analysis relies on state-of-the-art mathematical techniques together with a simple underlying model. 
In particular, we assume that particles move on an underlying two-dimensional lattice with stochastic dynamical rules and that self-propulsion only occurs in the left and right directions. 
These idealized features enable a rigorous hydrodynamic limit~\cite{kipnisScalingLimitsInteracting1998,erignouxHydrodynamicLimitActive2021}: when the lattice spacing tends to zero and the number of particles to infinity, the particle densities obey deterministic continuum equations, which we derive exactly.  
The resulting system differs from NRCH: it has some similar features but also reveals interesting new behavior which we outline next.

In our model, active particles alone result in stationary, motility-induced phase separation (commonly referred to as MIPS \cite{catesMotilityInducedPhaseSeparation2015}), but adding an extra population of passive (diffusing) particles can induce patterns, including traveling clusters, where self-organized groups of active particles push their passive counterparts around the system
\cite{stenhammarActivityInducedPhaseSeparation2015,wysockiPropagatingInterfacesMixtures2016}, see also~\cite{mccandlish2012}. 
The hydrodynamic equation reveals an unusual type of phase diagram where the spinodal curve for the liquid-vapor transition protrudes through the binodal, signaling the onset of pattern formation. In this region we demonstrate the existence of asymmetric traveling patterns and patterns formed of counterpropagating clusters. 
For large systems, these patterns feature sharp interfaces that separate dense liquid and dilute vapor regions. 
By studying the hydrodynamic equation in large domains, we relate traveling interfaces to those of static MIPS clusters, providing a new link between the equilibrium-like physics of phase separation and the dynamical patterns characteristic of nonreciprocity.

Our study provides an explicit and important example of how  nonreciprocal effective interactions can emerge at the macroscale from simple interactions at the microscale. Specifically, while the direct interactions among particles are reciprocal -- resulting solely from volume exclusion -- particle simulations and analyses of the hydrodynamic equation clearly demonstrate that the model exhibits the principles of nonreciprocal self-organization.
Furthermore, its idealized features allow us to draw sharp conclusions about large length and time scales, which would be extremely challenging to achieve from numerical simulations of more complex models. Our analysis of the large-system limit demonstrates how concepts from equilibrium phases and their interfaces can be applied to pattern-forming states. These features also facilitate a fully nonlinear treatment of the pattern-forming (traveling) steady states, revealing much more intricate behavior than could be predicted by linear stability analysis of the homogeneous state.

Based on these exact results, we identify several phenomena that should be considered as reference points for future studies of nonreciprocal systems.  Specifically: the protrusion of a spinodal line through the binodal is a generic mechanism for instability of static phase separation, leading to dynamical patterns; in large systems, such patterns may incorporate interfaces between dense and dilute regions resembling those found in static phase separation; the resulting patterns may travel with a fixed velocity or they may include different traveling objects that interact repeatedly.  These results significantly expand the phenomenology of nonreciprocally coupled mixtures.  The separation of length scales between sharp interfaces and macroscopic dense/dilute domains suggests focusing on dynamical patterns in a suitable large-system limit in order to draw useful analogies with static (thermodynamics) phase transitions.  We discuss these points further in later sections.

\section*{Results}
\subsection{Active-Passive Lattice Gas (APLG) Model}
\begin{figure*}[tb]
 \centering
 \includegraphics[width=1.0\linewidth]{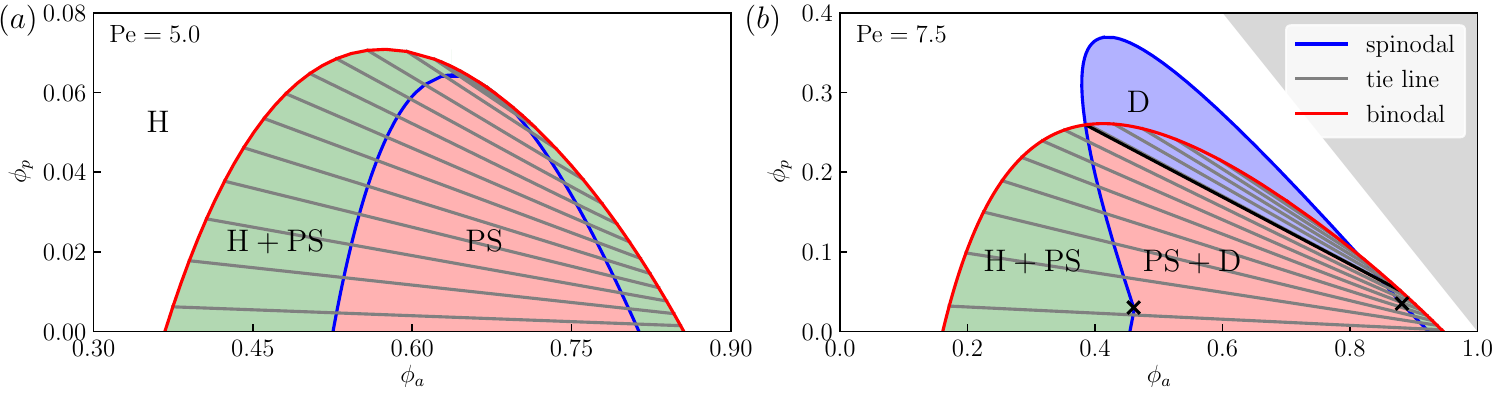}
 \caption{\textbf{Phase diagrams of the active-passive mixture for two values of $\pe$.} We show the diagram spanned by the active and passive particles' volume fractions $\phi_a$ and $\phi_p$, respectively, for (a) $\pe = 5$ and (b) $\pe = 7.5$. 
 The spinodal (blue curve) encloses the region of linear stability of homogeneous solutions. The binodal (red curve) encloses the region of phase separation. Intersection points of the tie lines (gray lines) with the binodal indicate the composition of the liquid and vapor phases.
Black crosses indicate bifurcations of co-dimension two; see text for discussion.
Regions with stable Homogeneous (H), Phase Separated (PS), and Dynamic (D) steady states are marked on the diagram. [In (b), the white region is left unlabelled because the boundary between H+D and H is unknown and the grey region corresponds to $\phi_a+\phi_p>1$, which is inaccessible due to size exclusion. The black tie line intersects the spinodal, so phase-separated states above this line are linearly unstable.]
}
\label{fig:phase}
\end{figure*}

We consider an active-passive lattice gas (APLG) model that extends the active lattice gases 
of~\cite{kourbane-housseneExactHydrodynamicDescription2018,erignouxHydrodynamicLimitActive2021,masonExactHydrodynamicsOnset2023a}.  It is defined on a two-dimensional periodic square lattice with spacing $h$. Placing at most one particle per site, we populate the lattice with three types of particles $\sigma \in \{+1,-1,0\}$: active particles oriented right ($\sigma = +1$), active particles oriented left ($\sigma = -1$) and passive particles ($\sigma = 0$).
Each lattice site $(i,j)$ has position $\x = (ih,jh)\in [0,\ell_x) \times [0,\ell_y)$.
The model dynamics can be split into four parts: (i)
passive particles attempt nearest neighbour jumps with jump rate $\Dx/h^2$ per adjacent site, where $\Dx$ is the spatial diffusion constant; 
(ii) active particles perform nearest neighbor random walk, weakly biased in the direction of their orientation to account for self-propulsion. In particular, a jump in the $\u$ direction (where $\vert \u \vert=h $) is attempted at rate $\Dx/h^2  + \frac{\vo}{2h^2}  ( \u \cdot \e_\sigma)$, where  $\e_\sigma = (\sigma, 0)$ is the particle's orientation, and $\vo$ is the self-propulsion speed; 
(iii) both types of particles are under an exclusion rule: if the target site of a jump is occupied, the jump is aborted; if the site is otherwise empty, the jump is executed; and (iv)
each active particle orientation flips at rate $\Dt$. The total numbers of active and passive particles are specified via their volume fractions $\phi_a$ and $\phi_p$, respectively, and the overall volume fraction is $\phi = \phi_a + \phi_p \in [0,1]$.

\paragraph{Hydrodynamic Limit}  

The number of lattice sites in the APLG is $\ell_x\ell_y/h^2$, and we identify the lattice spacing $h$ with the size of a particle.  
The hydrodynamic limit is $h\to0$ at fixed $\ell_x,\ell_y,\phi_a,\phi_p$: it corresponds to ``zooming out'', in order to describe motion on scales much larger than the particle size.  The APLG dynamics above has nontrivial $h$-dependence  [for example, the hopping rate is $O(h^{-2})$ but the orientation flip rate is $O(1)$].  This choice of scaling limit enables rigorous mathematical calculations~\cite{erignouxHydrodynamicLimitActive2021}. Other types of large-system limit are discussed in Appendix~\ref{sec:scaling}.

To take the hydrodynamic limit, it is convenient to rescale time and space by $\Dt^{-1}$ and $\sqrt{\Dx/\Dt}$ respectively, and introduce the P\'eclet number $ \pe = \vo / \sqrt{\Dx \Dt}$.
A configuration of the APLG is defined in terms of occupancies: $\eta_\sigma(\x,t) \in \{0,1\}$ is the number of particles of type $\sigma$ at site $\x$ and time $t$.
The hydrodynamic limit equations describe the evolution of the local densities $\rho_\sigma(\x,t)$ of particles of type $\sigma \in \{+1,-1,0\}$,
as the lattice spacing $h \to 0$. Building on \cite{erignouxHydrodynamicLimitActive2021}, we rigorously derive the hydrodynamic limit of the APLG model, obtaining exact macroscopic evolution equations for the densities $\rho_\sigma$ (see Appendix~\ref{sec:hydro}) 
	\begin{multline}\label{equ:main}
		\pt \rho_\sigma =  \div \left[ \ds( \rho ) \nab \rho_\sigma + \rho_\sigma \DD (\rho ) \nab \rho \right]  \\
	 - \pe \px \left[ \rho_\sigma s(\rho ) \pol  +\sigma \ds ( \rho ) \rho_\sigma \right]  - \sigma \pol,
	\end{multline}
with periodic boundary conditions. Here $\nabla = (\partial_x, \partial_y)$,  $\pol =  \rho_+ - \rho_-$ is the magnetisation, we also define
$\rho_a =\rho_+ + \rho_-$ as the active particle density, so
$\rho = \rho_a +\pas$ is the total particle density.
Further, 
$\ds(\rho)$ is the self-diffusion coefficient of a simple symmetric exclusion process~\cite{quastelDiffusionColorSimple1992,erignouxHydrodynamicLimitActive2021}
 and
\begin{align}
	\DD (\rho ) = [1- \ds (\rho)]/\rho, \quad s(\rho) = \DD(\rho ) - 1.
	\label{equ:DD}
\end{align}
Without activity ($\pe = 0)$, the APLG model reduces to a three-species symmetric simple exclusion process (SSEP), so it is natural that  $d_s$ appears in its hydrodynamic limit~\cite{quastelDiffusionColorSimple1992}.
Technically, the APLG model is of nongradient type in the sense of \cite{kipnisScalingLimitsInteracting1998}; this means that while \eqref{equ:main} is exact, there is no explicit expression for $d_s(\rho)$, although variational formulae are available~\cite{arita2017variational,spohn1990tracer}.
Even so, $d_s$ may be approximated to arbitrary accuracy, either numerically or analytically.  We use the polynomial approximation~\cite{masonMacroscopicBehaviourTwoSpecies2023}:
\begin{equation}
\label{eq_ds_approx}
	\ds (\rho) \approx (1-\rho)\Big( 1 - \alpha \rho + \frac{\alpha(2\alpha-1)}{2\alpha +1} \rho^2 \Big),
\end{equation}
with $\alpha = \pi /2 -1$, which exactly matches the value and the first derivative of $d_s(\rho)$ at $\rho = 0, 1$, and approximates $d_s$ very accurately for all $\rho$. For other accurate approximations, see \cite{illienNonequilibriumFluctuationsEnhanced2018,rizkallahMicroscopicTheoryDiffusion2022,aritaBulkDiffusionKinetically2018,brunaDiffusionMultipleSpecies2012,dabaghiTensorApproximationSelfdiffusion2023,nakazatoSiteBlockingEffect1980}. Here, we use \eqref{eq_ds_approx} in the numerical analysis of the exact hydrodynamic limit \eqref{equ:main}. 
Corrections to the mean-field approximation $\langle \eta_\sigma(\x,t)\eta_\sigma(\hat \x,t) \rangle \approx  \rho_\sigma(\x,t)\rho_\sigma(\hat \x,t)$ are important in the APLG model and enter the hydrodynamic limit; the mean-field approximation would lead to $d_s(\rho) = (1-\rho)$. The method to obtain \eqref{equ:main} is valid for any dimension greater than one.

Mathematical analysis of \eqref{equ:main} is challenging due to the density dependence of the coefficients.  However, since the active self-propulsion is only in the horizontal ($x$) direction, the equations are diffusive in the vertical direction~\cite{quastelDiffusionColorSimple1992}, so that any variation with respect to $y$ will converge to zero over time, and instabilities of the homogeneous state are also independent of $y$.  We exploit this symmetry throughout, restricting solutions of \eqref{equ:main} to the form $\rho_\sigma (x,y,t) = \rho_\sigma (x,t)$. 
Four dimensionless parameters govern these solutions: the 
P\'eclet number $\pe$, the rescaled domain length in the horizontal axis $L = \ell_x\sqrt{\Dt/\Dx}$, the active volume fraction $\phi_a$, and passive volume fraction $\phi_p$.

\subsection{Overview of phase behavior}
\begin{figure*}
\centering
\includegraphics[width=\textwidth]{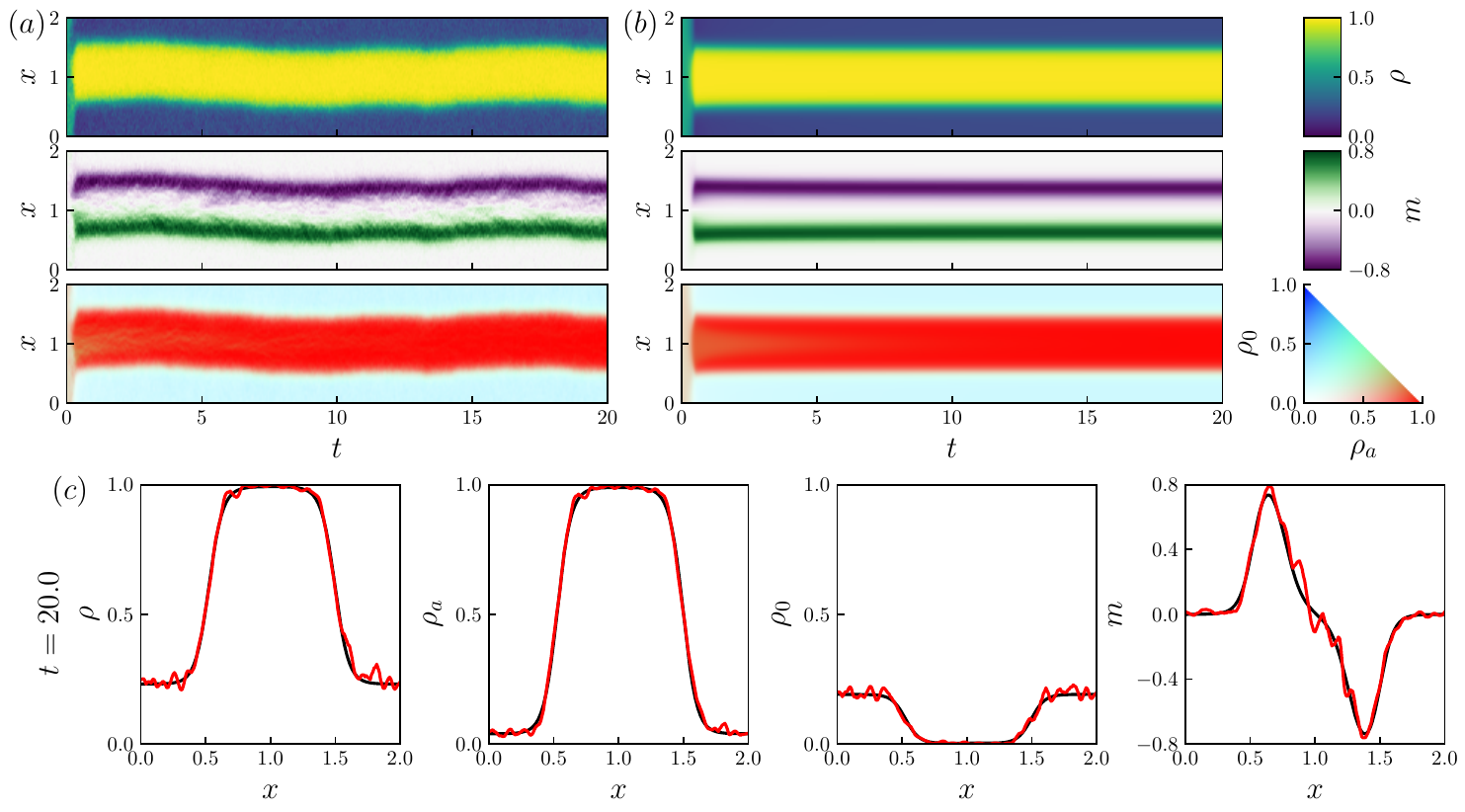}
\caption{\textbf{Phase-separated stationary solution.}
(a),(b) KymographsAppendix showing the spatiotemporal dynamics of a phase-separating solution. (a) $y$-averaged local density of a particle simulation (see \methodsref). (b) Numerical solution to \eqref{equ:main}, the initial condition is a homogenous state with a uniform random perturbation (see Appendix~\ref{sec:timedep} for details, Eq.~\eqref{rand_ic}). (c) Density profile at $t=20$. Black lines display the solution to \eqref{equ:main}, and red lines display the local density of the particle simulation. Parameters: $\pe = 20$, $L =2$, $\phi_a = 0.5$, $\phi_p = 0.1$, $\Delta x = 0.01$ (PDE model), $h= 0.01$ (particle model).
}
\label{fig:MIPS}
\end{figure*}

\begin{figure*}
\centering
\includegraphics[width=17.8cm]{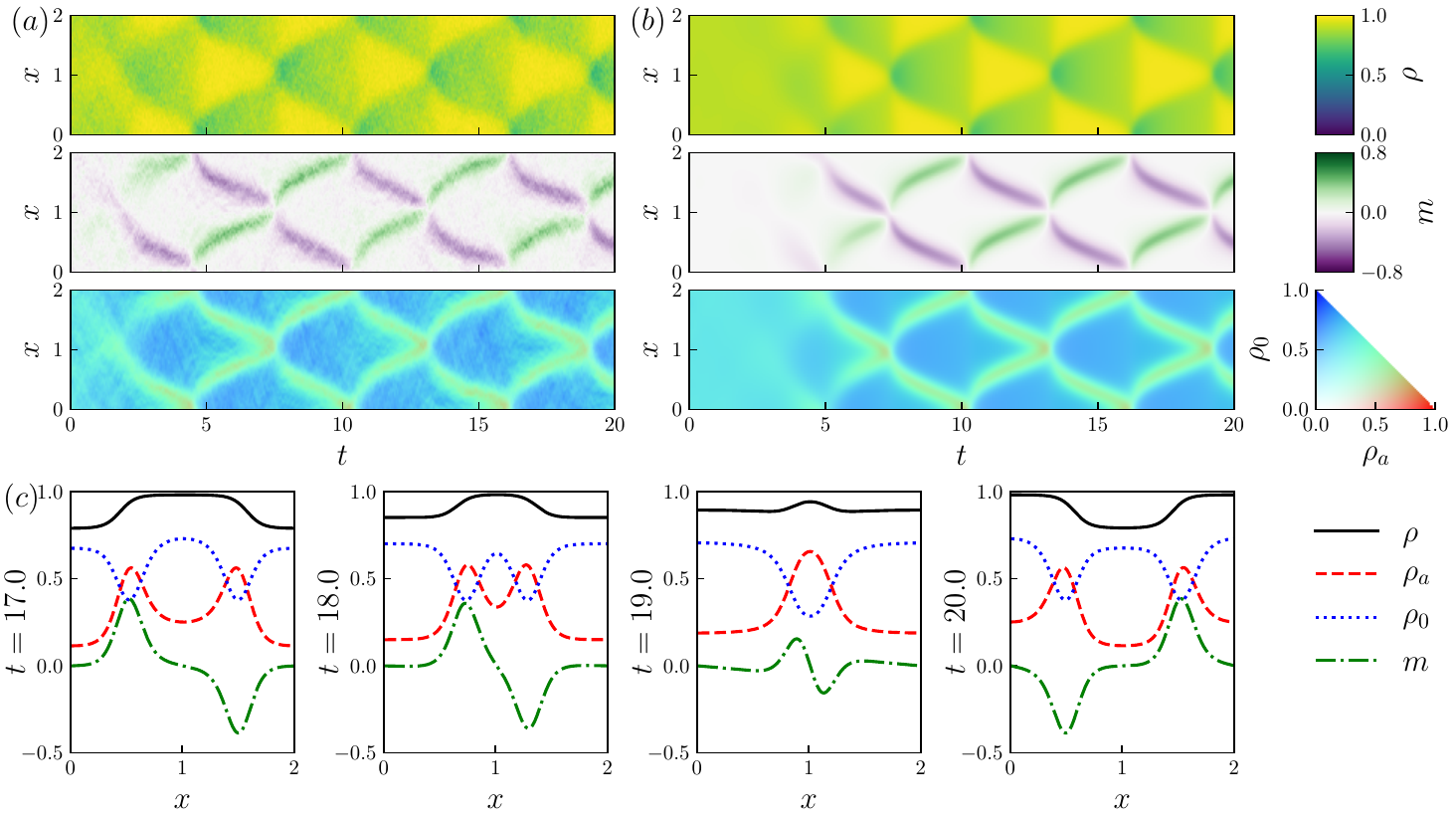}
\caption{\textbf{Counter-propagating solution}
(a),(b) Kymographs showing the spatiotemporal dynamics of a CP solution. (a) The $y$-averaged local density of a particle simulation (see \methodsref). (b)~The solution to \eqref{equ:main} whose initial condition is a homogenous state with a uniform random perturbation (see Appendix~\ref{sec:timedep}, Eq.~\eqref{rand_ic}). (c) Density profiles before, during, and after the collision of two counter-propagating interfaces ($t=17,18,19,20$). 
Parameters: $\pe = 20$, $L =2$, $\phi_a = 0.3$, $\phi_p = 0.6$, $\Delta x = 0.01$, $h= 0.01$.
}
\label{fig:CP}
\end{figure*}

The APLG model supports different dynamical phases.  We focus on behavior in the thermodynamic limit of large domains ($L \gg 1$), which allows some properties to be derived analytically.  
As in pure active systems~\cite{masonExactHydrodynamicsOnset2023a}, the APLG supports phase-separated (PS) stationary states with macroscopic dense (liquid) and dilute (vapor) regions.
The densities of active and passive particles in the liquid and vapor phases trace out the binodal curve in the ($\phi_a,\phi_p)$ plane, see Fig.~\ref{fig:phase}(a).
For $L\gg1$, this curve can be calculated numerically exactly following the prescription of~\cite{solonGeneralizedThermodynamicsPhase2018}. This relies on the observation that, for PS states, local concentrations of passive particles and vacancies are proportional 
\begin{equation}
\rho_0(x) = \nu [ 1- \rho(x) ]
\label{equ:nu-local}
\end{equation}
(see \methodsref), with
\begin{equation} \label{equ:nu}
\nu = \phi_p/(1-\phi).
\end{equation}  

We also compute the spinodal curve as the limit of stability of the homogeneous state (which is $\rho_\pm=\phi_a/2$, $\rho_0=\phi_p$).
Linear stability analysis of \eqref{equ:main} shows that the homogeneous state is unstable when $\pe$ is sufficiently large and $\phi_p$ is sufficiently small (see \methodsref). For the case shown in Fig.~\ref{fig:phase}(a), the dominant eigenvalue of the stability problem is always real, as in the pure active case.  Then, the binodal and spinodal are tangent at the critical point (see Fig.~\ref{fig:phase}(a)), which is the standard phenomenology of liquid-vapor phase coexistence for two-component mixtures. Inside the spinodal, the homogeneous state is linearly unstable, and the phase-separated (PS) state is stable; between the spinodal and binodal, the PS state is globally stable while the homogeneous state is metastable.

For larger $\pe$, this familiar scenario changes qualitatively. In particular, the dominant eigenvalues of the linear stability problem may become complex, leading to dynamical steady states.  Such a scenario is illustrated in  Fig.~\ref{fig:phase}(b).  Crosses on the spinodal mark co-dimension two (Bogdanov--Takens) bifurcations~\cite{kuznetsovBook} where the dominant eigenvalue becomes complex and the instability of the homogeneous state changes from a pitchfork to a Hopf bifurcation~\cite{crossPatternFormationOutside1993}. Crucially, the binodal curve can still be computed for this system: we observe that the spinodal protrudes through the binodal. 
This effect is intrinsically linked to the existence of nonreciprocal effective interactions between active and passive densities and signals the onset of new physics \cite{youNonreciprocityGenericRoute2020, wysockiPropagatingInterfacesMixtures2016, wittkowskiNonequilibriumDynamicsMixtures2017}.  (See Appendix~\ref{sec:linear_stab} for details of the stability analysis.)

Within the protruding part of the spinodal, the homogeneous state is linearly unstable, and PS states do not exist. It follows that some dynamical steady states must be present in this region. Moreover, in regions where  PS states do exist, it may be that (at least) one of the coexisting phases is linearly unstable, which also renders the PS state unstable.  The result is that dynamical steady states must exist throughout the blue-shaded region D in Fig.~\ref{fig:phase}(b).   Note that this does not rule out the existence of dynamical states elsewhere. Also, while the dominant eigenvalue for the instability is complex for a large part of the spinodal in Fig.~\ref{fig:phase}(b), the resulting steady state may still be (stationary) PS, showing that steady-state properties cannot be deduced directly from linear stability analysis. Dynamical behavior similar to the APLG also occurs in other nonreciprocal systems~\cite{youNonreciprocityGenericRoute2020, wysockiPropagatingInterfacesMixtures2016, wittkowskiNonequilibriumDynamicsMixtures2017}: we will see below that the APLG provides a specific microscopic model where such generic phenomena can be analyzed quantitatively, including in the thermodynamic limit of large system size.

We emphasize that the spinodal marks the stability limit of homogeneous states, but PS stationary states can also exhibit other linear instabilities corresponding to critical exceptional points or exceptional phase transitions~\cite{fruchartNonRecip2021,youNonreciprocityGenericRoute2020,suchanekTimereversalParitytimeSymmetry2023}. In such transitions, broken translational symmetry of the PS state plays an important role. We expect the (blue) region of purely dynamical steady states in Fig.~\ref{fig:phase}(b) to extend to lower $\phi_p$.  We discuss such cases below, as well as other state points where both static and dynamical states are linearly stable.

\subsection{Illustrative steady states}
\label{sec:illust}

We present numerical results that illustrate the behavior of the APLG, including direct simulation of the particle model by the Gillespie algorithm \cite{erbanStochasticModellingReaction2020}, and time-stepping the (deterministic) partial differential equation~\eqref{equ:main} (see Methods).  We consider fairly small domains so that the particle model simulations are tractable. [The total number of particles is approximately $(\ell_y/\ell_x) \phi (L/h)^2 $, such that the simulations of Figs. \ref{fig:MIPS} and \ref{fig:CP} involve thousands of particles.]

\paragraph{Phase-separated (PS) solutions} 
The binodal construction (Fig.~\ref{fig:phase}) demonstrates the existence of stationary solutions to \eqref{equ:main} for large system size $L$. These consist of large $O(L)$ liquid and vapor regions, separated by interfaces of $O(1)$ width. Such phase-separated states are familiar in NRCH, when the nonreciprocity parameter is not too large.  In the current setting, the phase separation is caused by the particles' self-propulsion together with their excluded volume interactions: this is an example of motility-induced phase separation (MIPS)~\cite{catesMotilityInducedPhaseSeparation2015,filyAthermalPhaseSeparation2012,catesWhenAreActive2013,brunaPhaseSeparationSystems2022}, which has been previously characterized in the pure active case ($\phi_p=0$) of this model~\cite{kourbane-housseneExactHydrodynamicDescription2018}.

Fig.~\ref{fig:MIPS} shows results of particle-based simulations of the APLG and corresponding numerical solutions of  \eqref{equ:main} in a domain of size $L=2$.  Phase separation occurs in both cases, starting from homogeneous initial states.  
We find that the dense phase is dominated by active particles, while the dilute phase is mostly passive. 
In fact, denser phases always have lower concentrations of passive particles because of 
\eqref{equ:nu-local}.
As usual for MIPS, the magnetization $m$ is large in the interfacial regions but very small within the phases. Note that while the analytic binodal computation shows that stationary PS states exist, these numerical results also show that they are stable (for our choice of parameters).

\begin{figure*}
\centering
\includegraphics[width=17.8cm]{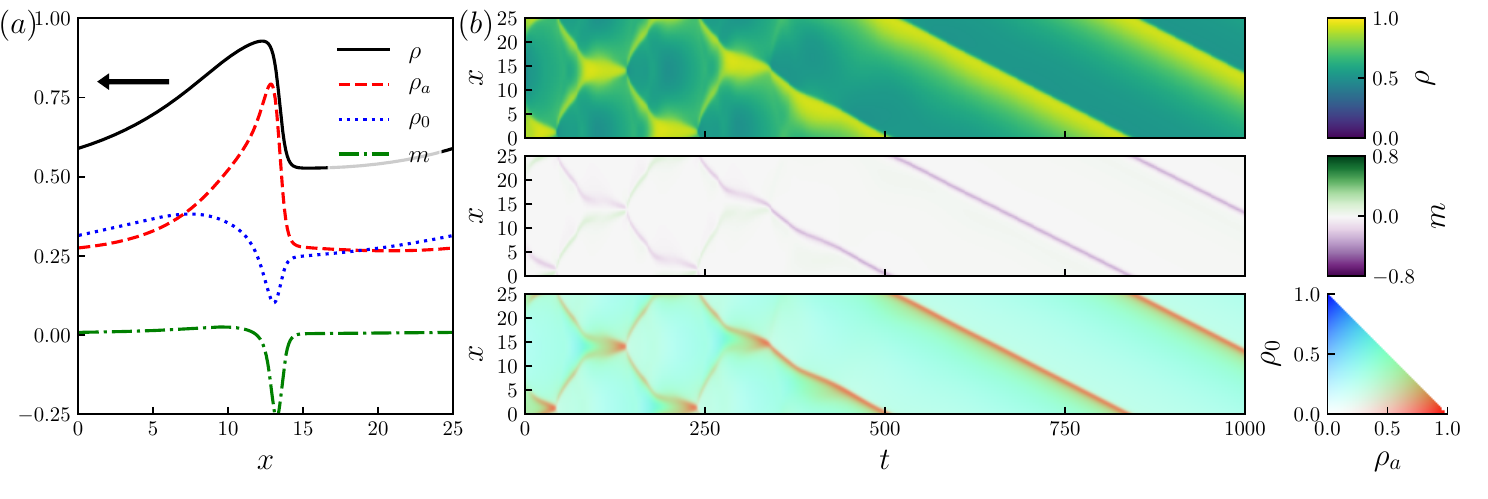}
\caption{\textbf{Traveling solution.} Traveling profile solution to \eqref{equ:main}, whose initial condition is the homogeneous state perturbed by a combination of left and right traveling unstable sinusoidal modes and a uniform random perturbation (see Appendix~\ref{sec:timedep}, Eqs.~(\ref{rand_ic},\ref{left_right_ic}) for further details).
(a) Density profile at $t=500$. 
(b) Kymographs showing the spatiotemporal dynamics. 
Parameters: $\pe = 7.5$, $L =25$, $\phi_a = 0.36$, $\phi_p = 0.3$, $\Delta x = 0.05$. Speed of propagation in \eqref{equ:travel-simple} is  $c = -1.8832$.}
\label{fig:TP}
\end{figure*}

\paragraph{Counter-propagating (CP) and Traveling (T) solutions}
We now turn to dynamical steady states, which appear in the (blue) region D in Fig.~\ref{fig:phase}(b).  We focus on two types:  CP solutions retain an overall left-right symmetry with clusters of particles moving in both directions; T solutions break left-right symmetry, leading to a density profile that travels at a fixed speed. Specifically
\begin{equation}
\label{equ:travel-simple}
\rho_\sigma(x,t)=\varrho_\sigma(x-ct/L),
\end{equation}
where $c$ is a constant so the wave velocity is $c/L$;  the reason for this $L$-dependence will be discussed below. 
 (There is an analogy of CP and T solutions with standing waves and traveling waves respectively, but note that both CP and T solutions are strongly anharmonic.)

An example CP state is shown in Fig.~\ref{fig:CP}, which again compares direct simulation of the particle model with the numerical solution of~\eqref{equ:main} 
whose initial condition is a homogenous state with a uniform random perturbation.
This perturbation grows via an instability that involves counterpropagating sinusoidal waves with equal speeds and growth rates. After this transient growth, the resulting time-periodic state consists of two oppositely polarised active clusters that move in opposite directions.  As they move, they accumulate passive particles in front of them via a ``snowplow effect''. The clusters collide, they pass through each other, and the cycle continues (see Appendix~\ref{sec:diagrams} for additional discussion). In the example of Fig.~\ref{fig:CP}, clusters appear to accelerate during collisions; a simpler example of this effect was observed in \cite{ralph2020one}.

Time-periodic states similar to CP have been observed in other active and nonreciprocal systems~\cite{agranovThermodynamicallyConsistentFlocking2024,frohoff-hulsmannSuppressionCoarseningEmergence2021,frohoffResonance2023,frohoffUniversal2023}: their presence here emphasizes that they are generic.  
In particular, the APLG still supports the rich dynamical phenomenology of other active-passive mixtures, despite its idealized features.  
The conection between the APLG and other nonreciprocal systems is discussed further in Appendix~\ref{sec:nonreciprocity}.
Despite the analogy with standing waves, we emphasize that these CP clusters experience complex (nonlinear) scattering processes when they meet (see, for example, Fig.~\ref{fig:CP_large}).

An illustrative T state is shown in Fig.~\ref{fig:TP}, obtained by time-stepping \eqref{equ:main} for a larger system ($L=25$).
After an initial transient which resembles a CP solution, we find a solution of the form of \eqref{equ:travel-simple}.
As in the CP case, this consists of a localized packet of active particles that pushes passive particles in front of it (see Appendix~\ref{sec:diagrams}).  We find numerically that such solutions can have a variety of shapes and nontrivial dependence on system size.
To simplify this diverse behavior, we again turn to large systems ($L\gg1$), which enables analytic progress, by analogy with PS states.  Hence we reveal new connections between dynamical patterns in this system and equilibrium phase coexistence.

\begin{figure*}
\centering
\includegraphics[width=\textwidth]{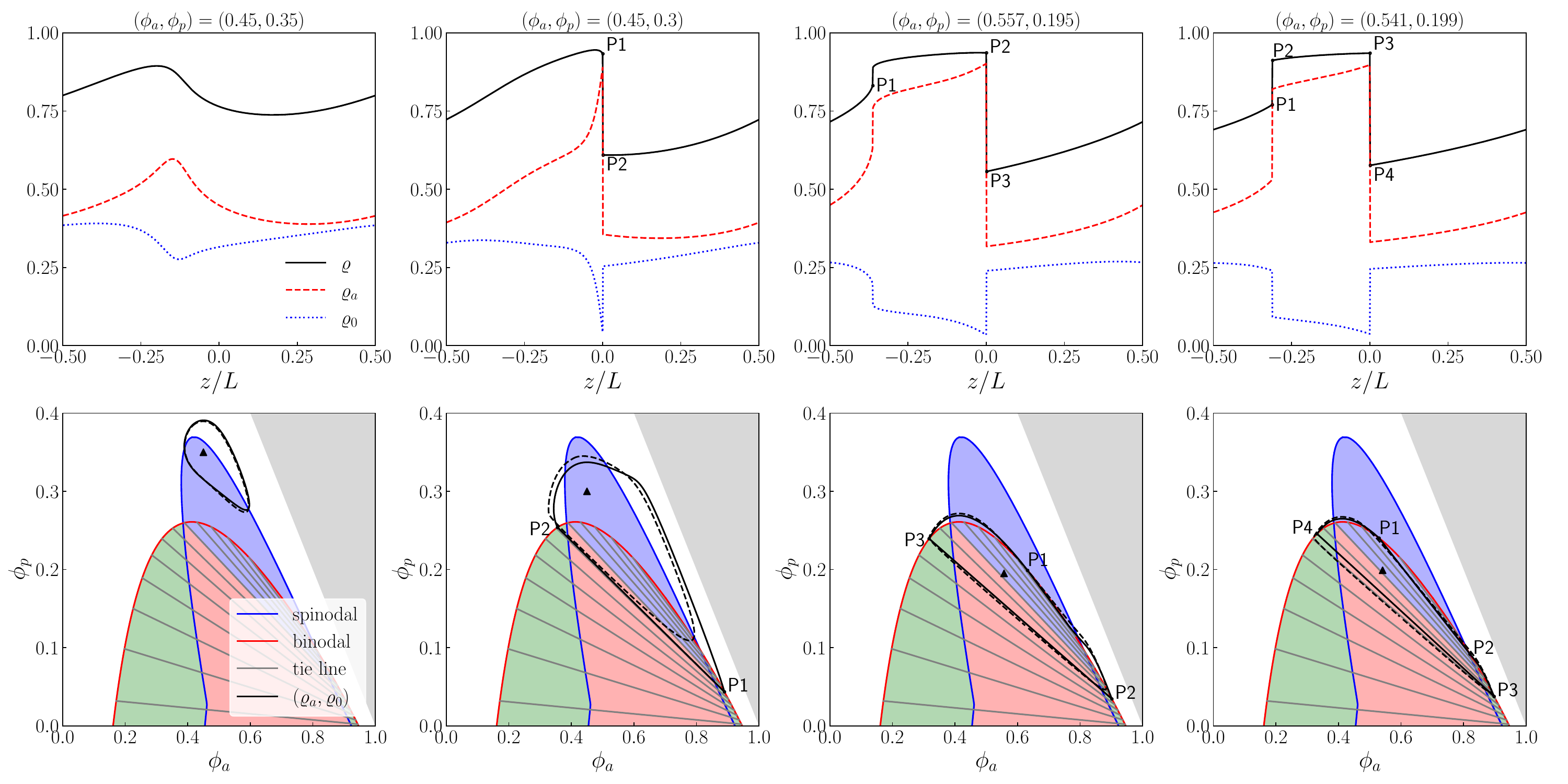}
\caption{\textbf{Traveling solutions in the large system-size limit.}
T solutions of the finite-$L$ problem \eqref{equ:travelling} and the large-$L$ problem \eqref{equ:outer_leading} at $\pe=7.5$.
(Top) Solutions to  \eqref{equ:outer_leading} as a function of $z/L$.
(Bottom)~Solutions plotted parametrically on the phase diagram of Fig.~\ref{fig:phase}(b), with dashed lines for solutions of \eqref{equ:travelling} and solid lines for solutions of \eqref{equ:outer_leading}. Volume fractions $(\phi_a,\phi_p)$ for each column are given in the top, also marked as black triangles in the bottom row.
Other parameters:
$N=1024$ for numerical solution of  \eqref{equ:outer_leading} and $L=25$,  $N = 1600$ for \eqref{equ:travelling}.  Wave speeds for (a,b,c,d) are $c=1.95, 1.39,0.41, 0.31$ (2 decimal places). 
}
\label{fig:outer}
\end{figure*}

\subsection{Traveling solutions in large systems} \label{sec:travelling}

Recall that $L\gg 1$ corresponds to $\ell_x\gg \sqrt{D_T/D_R}$: this means that the system size is much larger than the intrinsic length scale associated with particle motion.
Some active-matter and nonreciprocal models \cite{youNonreciprocityGenericRoute2020,solon2013revisiting} support traveling phase-separated states where a large system has macroscopic liquid and vapor domains separated by sharp interfaces, which move at constant velocity.  However, this is not possible in the APLG.   
To see this, substitute \eqref{equ:travel-simple} into \eqref{equ:main} and write $z=x-ct/L$ to obtain
\begin{multline}\label{equ:travelling}
-(c/L) \varrho_\sigma' =  \partial_{z}\left[ \ds( \varrho ) \varrho_\sigma' + \varrho_\sigma \DD (\varrho )  \varrho' \right]  \\
	 - \pe \partial_{z} \left[ \varrho_\sigma s(\varrho ) \poltravel  +\sigma \ds ( \varrho ) \varrho_\sigma \right]  - \sigma \poltravel,
\end{multline}
where $\varrho_\sigma$ and $\poltravel = \varrho_+ - \varrho_-$ denote the densities and magnetization in the traveling frame [recall \eqref{equ:travel-simple}], and primes indicate derivatives with respect to $z$. 
Within the bulk of the phases, $\varrho'=0$ so $\poltravel=0$ in the bulk.  Summing \eqref{equ:travelling} over $\sigma$ to obtain the total density $\varrho$ and integrating then yields $(c/L)\varrho + \varrho' - \pe (1-\varrho)\poltravel = J$ where $J$ is an integration constant.  Evaluating this expression inside the two phases where $\poltravel=0=\varrho'$, one finds $\varrho=L J/c$ so both phases would need to have equal densities, ruling out any traveling phase-separated states (the special case $c=0=J$ recovers the PS state).
This exact analysis illustrates the value of our exact hydrodynamic description: it is very difficult to extrapolate such results for large systems based on numerical simulations alone, especially because time-stepping Eq.~\eqref{equ:main} is expensive in large domains.

There are macroscopically smooth T solutions of \eqref{equ:travelling} satisfying $\varrho'_\sigma=O(1/L)$ [with $c=O(1)$ and $\poltravel=O(1/L)$].  
We do find such solutions numerically (see Methods). However, Fig.~\ref{fig:TP}(a) includes an interfacial region where $\varrho$ varies quickly in space, hinting at the existence of solutions with traveling narrow interfaces [$\varrho'=O(1)$]. 
\begin{figure}[b]
\centering
\includegraphics[width=1.0\columnwidth]{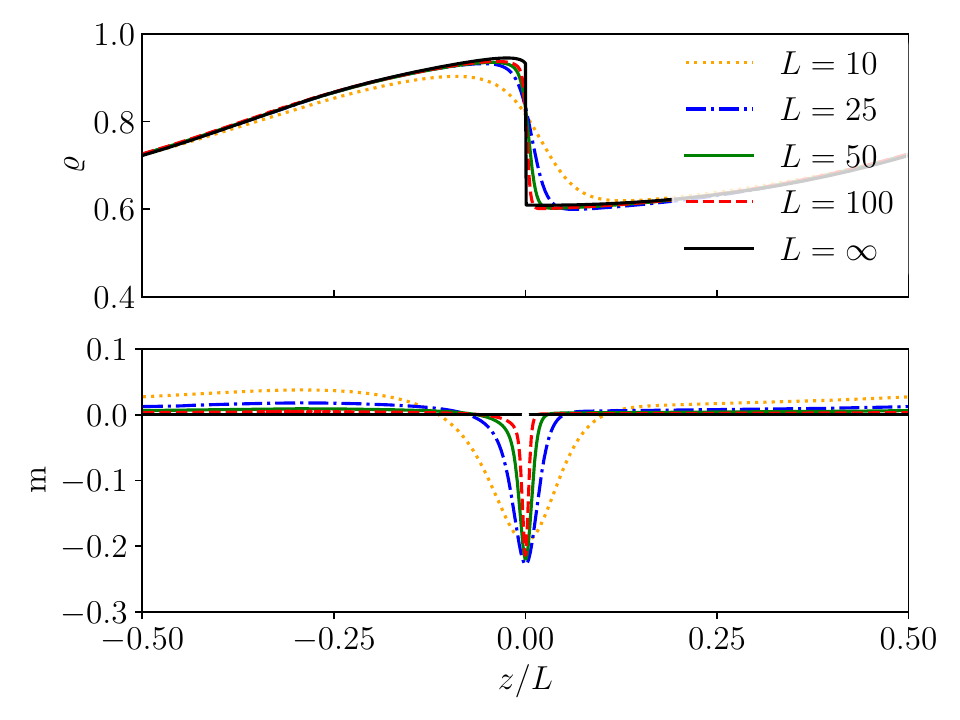}
\caption{\textbf{Domain-size dependence of traveling solution.} T solutions of the finite-$L$ problem \eqref{equ:travelling} for different values of $L$ and the large-$L$ problem \eqref{equ:outer_leading} at $\pe=7.5$. Total density $\varrho$ (top) and magnetisation $\poltravel$ (bottom). The magnetisation for $L=\infty$ is $\poltravel = 0$ for $|z|>0$.
Parameters as in Fig.~\ref{fig:outer}(b).
 }
\label{fig:convergence}
\end{figure} 

\begin{figure*}
\centering
\includegraphics[width=1.0\linewidth]{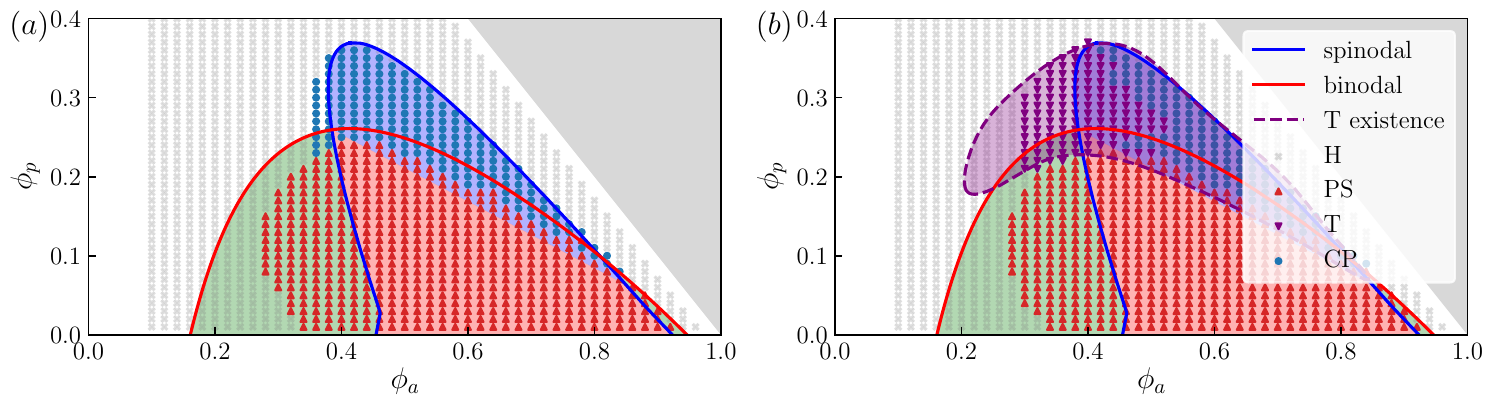}
\caption{\textbf{Stability of T and CP solutions.} Phase diagrams from Fig.~\ref{fig:phase}(b), overlayed with the types of long-time solutions of \eqref{equ:main}: homogeneous (H), phase-separated (PS), traveling (T), and counter-propagating (CP), see \methodsref\ for the classification. 
The initial condition is the homogeneous state,
perturbed by 
(a) left and right traveling unstable sinusoidal modes
(b) left traveling unstable sinusoidal modes (see Appendix~\ref{sec:timedep} for details).
Parameters: $\pe = 7.5$, $L =25.0$, $\phi_a = 0.36$, $\phi_p = 0.3$, $\Delta x = 0.05$.
 }
\label{fig:steady_states_finite}
\end{figure*}

These T states with sharp interfaces do indeed exist.  We find them systematically using the method of matched asymptotic expansions \cite{holmes2012introduction}, with results shown in Fig.~\ref{fig:outer}. This method consists of expanding solutions in an asymptotic series in the small parameter $1/L$: this is a controlled approximation scheme that provides accurate results for large systems. Specifically, we seek a solution to \eqref{equ:travelling} with an interface at $z=0$, and
 we split the domain of $z$ into two overlapping subdomains, 
an inner region around the wave front where $|z| = O(1)$ and an outer region far from the front  ($|z| \gg 1$). 
Details of the calculation are given in \methodsref.
In the inner region, the problem reduces, at leading-order in $1/L$, to the same Eqs.~(\ref{equ:tie_1},\ref{equ:tie_1_rho0},\ref{equ:tie_1_m}) that govern PS solutions. That is, the leading order of sharp interfaces in T states coincide with liquid-vapor interfaces in PS, and connect points on the binodal by tie-lines (recall Fig.~\ref{fig:phase}).
In the outer region, we can eliminate $\poltravel$, and the system reduces to two equations 
\begin{subequations}\label{equ:outer_leading}
\begin{align}
	-(c/L)  \varrho_a' =\ & \partial_{ z}\left[ \ds(  \varrho )  \varrho_a' +  \varrho_a \DD ( \varrho )   \varrho' \right]  \\
	 & + \frac{\pe^2}{2}  \partial_{ z} \left \{ \left[  \varrho_a s(  \varrho ) +  \ds (  \varrho ) \right] \partial_{ z} \left[ \ds ( \varrho)  \varrho_a \right]  \right \}, \nonumber
\\ 	
	 	-(c/L)  \varrho_0' =\ & \partial_{ z}\left[ \ds(  \varrho )  \varrho_0' +  \varrho_0 \DD ( \varrho )   \varrho' \right]  \\
&	 + \frac{\pe^2}{2}  \partial_{ z} \left \{  \varrho_0 s(  \varrho )  \partial_{ z} \left[ \ds ( \varrho)  \varrho_a \right]  \right \}, \nonumber
\end{align}
\end{subequations}
where $\varrho_a = \varrho_+ + \varrho_-$. Eq.~\eqref{equ:outer_leading} is to be solved with periodic boundary conditions at $z= \pm L/2$ and matching conditions to the binodal densities as $z\to 0^\pm$. 

We numerically solve \eqref{equ:outer_leading} for the densities $\varrho_a, \varrho_0$ and the speed $c$ and show the results for different combinations of $(\phi_a, \phi_p)$ in Fig.~\ref{fig:outer}.
The top row of Fig.~\ref{fig:outer} shows four solutions of \eqref{equ:outer_leading}, and the bottom row shows the same solutions overlaid on the phase diagram, together with the corresponding solutions to the original finite-$L$ problem \eqref{equ:travelling}.
We see that the finite-$L$ solutions follow the tie lines, similar to the matched asymptotic solutions.  Fig.~\ref{fig:convergence} shows how direct solutions of \eqref{equ:travelling} approach the asymptotic solution as $L$ is increased, confirming that the matched asymptotic analysis provides accurate results. Further, the spinodal curves of the outer problem \eqref{equ:outer_leading} agree with the spinodals of the full problem (see Appendix~\ref{sec:approx_spi}).

It is in the outer region formulation \eqref{equ:outer_leading} that the effective nonreciprocal interactions become most clearly apparent in the hydrodynamic equation \eqref{equ:main}. In particular, we show in Appendix~\ref{sec:nonreciprocity} that \eqref{equ:outer_leading} can be expressed as a nonlinear cross-diffusion system for the active and passive densities $\rho_a$, $\rho_0$ with nonreciprocal effective interactions between them.

We identify different types of T solutions on varying the volume fractions $\phi_p, \phi_a$. 
For pairs $(\phi_a, \phi_p)$ in the upper part of region D, we observe smooth (periodic) T solutions with no inner region,
see Fig.~\ref{fig:outer}(a). 
Reducing $\phi_p$ in the phase diagram, we find solutions with a single interface (Fig.~\ref{fig:outer}(b)). The inner region occurs between points P1 and P2 and lies along a tie line between two points on the binodal.
Another interesting case is when the high-density part of the solution enters the binodal, leading to T solutions with two interfaces. 
Fig.~\ref{fig:outer}(c) shows this transition point: the first inner region is a single point (P1), tangent to the binodal at its critical value, and the second interface occurs between P2 and P3. Slightly reducing $\phi_a$, we obtain solutions with two interfaces (Fig.~\ref{fig:outer}(d)): these have two inner regions that both follow tie-lines.  

The array of solutions in Fig.~\ref{fig:outer} illustrates the rich phenomenology of the APLG in large domains.  In particular, the appearance of narrow interfaces in T solutions is a surprising feature, since it connects these dynamically-patterned states to the equilibrium-like constructions of the binodal and the associated interfacial profiles.  
While the analytical characterization of T solutions is not possible for CP solutions, numerical simulations show that  CP solutions also feature narrow interfaces that follow the tie lines (see Fig.~\ref{fig:CP_large}, in Appendix).
The distinction between narrow interfaces and macroscopically smooth profiles would be very challenging to characterize by direct numerical solution of \eqref{equ:main}: the method of matched asymptotics makes this possible.

\subsection{Multistability} \label{sec:multistab}

We have characterized T solutions, but this does not guarantee that the time-dependent system will converge to a T state, nor even that they are locally stable. There could also be other T solutions with different speeds $c$.
To address this question, Fig. \ref{fig:steady_states_finite} shows the types of some long-time solutions obtained by numerically time-stepping \eqref{equ:main}. The system is initialized with a zero-magnetization homogeneous state, perturbed by the most unstable eigenmode of the linear stability analysis. In the case of complex eigenvalues, solutions in Fig.~\ref{fig:steady_states_finite}(a) use a linear combination of left- and right-moving modes, while Fig.~\ref{fig:steady_states_finite}(b) uses only the left-moving mode.  These systems converge to steady states, which we characterize as H, PS, T, or CP (see \methodsref\ for further details). 

The results are consistent with Fig.~\ref{fig:phase} and demonstrate the existence and stability of T and CP states, for suitable parameters. 
Comparing Figs.~\ref{fig:steady_states_finite}(a,b), the steady state also depends on the initial conditions: initializing with a left-moving mode favors T solutions while symmetric initialization favors CP.  Fig.~\ref{fig:steady_states_finite}(b) also shows the range of parameters over which we were able to find T solutions via matched asymptotics, showing that nonsymmetric initialization may not be sufficient to drive the system into T states, even if they exist. Together, these observations demonstrate multiple dynamical attractors, as may be expected for such complex PDE systems. Fig.~\ref{fig:multistability} (in Appendix) shows an explicit example where both T and PS solutions exist for the same parameters.  Understanding the basins of attraction of different states in more detail is an important challenge for future work.

\section*{Discussion}
\label{sec:discuss}

We introduced the APLG as a microscopic model of interacting particles and characterized its hydrodynamic behavior.  In addition to PS states familiar from pure active systems, we find rich behavior, including that the spinodal curve can protrude through the binodal.  This signals the existence of dynamical steady states, which we classify according to their symmetries.  Some of these results are similar to previous work on the NRCH equation~\cite{sahaScalarActiveMixtures2020,youNonreciprocityGenericRoute2020,frohoff-hulsmannSuppressionCoarseningEmergence2021,braunsNonreciprocalPatternFormation2023}, but our hydrodynamic PDE includes the magnetization $m$ as a slow hydrodynamic variable, in addition to the two conserved densities $\rho_a,\rho_0$; it also has a distinct set of nonlinearities.    
Hence, our approach of deriving hydrodynamic equations exactly from a microscopic model complements the generic description of nonreciprocal systems via the NRCH equation.
As already noted in Sec.~\ref{sec:illust}, it is important that the APLG supports the complex behavior of other nonreciprocal models, despite the idealized modeling assumptions in its definition.

\paragraph{APLG phenomenology}

To tame the complexity of the APLG's behavior, we focussed on large domains $L\gg1$.  This enables numerically exact computation of the binodal and spinodal curves and precise characterization of traveling solutions that involve sharp interfaces. Surprisingly, interfacial profiles in static and traveling states both obey \eqref{equ:nu-local}, which also describes the tie-lines in the phase diagram.   If such connections are generic in nonreciprocal systems, they would have broad consequences for understanding their phase diagrams, including possibilities for long-ranged order similar to equilibrium.  We also demonstrated multistability: qualitatively different solutions can exist,  for the same parameters.

 Our results also raise further questions for the APLG, including the existence of dynamical states in large systems with patterns on finite length scales, and associated questions of wavelength selection \cite{braunsNonreciprocalPatternFormation2023,duan2023,frohoff-hulsmannSuppressionCoarseningEmergence2021,solon2015pattern}. While this work analyzed deterministic hydrodynamic equations, the theory of fluctuating hydrodynamics can also be developed for such models~\cite{kipnisScalingLimitsInteracting1998,agranov2021exact,martin2307transition}
 by retaining corrections to the limit $h\to0$. This enables studies of metastability and the role of noise in determining a system's eventual steady state.  There might also be interesting corrections to the large-$L$ behavior studied here.  One may also expect new behavior on replacing the two-state active-particle orientation ($\sigma=\pm$) with a continuous degree of freedom: hydrodynamic limits can be derived in this case~\cite{erignouxHydrodynamicLimitActive2021,masonExactHydrodynamicsOnset2023a} but the resulting equations are challenging to analyze.  Future work should investigate these issues.

\paragraph{Implications for nonreciprocal systems}

To provide a broader perspective, we note that the APLG model serves as an example of a general class of nonequilibrium mixtures.  Its exact hydrodynamic limit relies on specific modeling assumptions, including a nontrivial $h$-dependence of the microscopic rates.  
However, the emergent large-scale picture is generic, including the appearance of complex stability eigenvalues, as observed in the NRCH and other nonreciprocal systems.  
As usual in mixtures of active and passive particles~\cite{wysockiPropagatingInterfacesMixtures2016,wittkowskiNonequilibriumDynamicsMixtures2017,youNonreciprocityGenericRoute2020}, this is driven by nonreciprocal effective interactions that appear in their hydrodynamic descriptions, notwithstanding that the microscopic interactions between particles are reciprocal.

For the APLG, we find that the protrusion of the spinodal through the binodal introduces a novel mechanism by which nonreciprocal behavior can arise, and it would be interesting to explore additional examples of this phenomenon. This could be achieved by considering hydrodynamic limits for other microscopic models, such as mixtures of different types of active particles~\cite{kolb2020active}), or within the framework of field theories like the NRCH, independent of their underlying microscopic descriptions. For a recent example of an active system where the spinodal protrudes through the binodal, see \cite{spinney2024non}, see also~\cite{bertin2024bias}.

An important takeaway from this work is the tractability of the large-system limit.  Large systems are essential for understanding equilibrium phase behavior, and it is valuable to identify nonequilibrium situations where such systems can be analyzed precisely.
In the case of the APLG model, this is facilitated by the presence of domain walls, which appear both in static phase separation and in dynamical steady states.
The scale separation between interfacial width and system size allows this behavior to be tackled using the method of matched asymptotics, providing a new route for characterizing dynamical pattern-forming states, such as those observed numerically in \cite{wysockiPropagatingInterfacesMixtures2016,wittkowskiNonequilibriumDynamicsMixtures2017,kolb2020active}.  We look forward to future work in this direction, which holds promising opportunities to bridge theories of pattern formation with those of equilibrium phase behavior (see also Appendix~\ref{sec:nonreciprocity}).

\addtocounter{section}{1}
\stepcounter{section}
\section*{Methods}\label{sec:methods}
\paragraph{Coexisting Phases}
For large systems, $L\gg1$, stationary phase-separated (PS) states contain large domains of liquid and vapor phases. The phases' (total) volume fractions are denoted $\phi_l, \phi_v$, which we compute by
generalising the method of \cite{masonExactHydrodynamicsOnset2023a, solonGeneralizedThermodynamicsMotilityinduced2018, solonGeneralizedThermodynamicsPhase2018}.
Specifically, 
setting $\partial_t\rho=0=\partial_t\rho_0$ in \eqref{equ:main} and integrating yields
\begin{align}
	\label{equ:tie_1}
		J &= \p_x \rho - \pe (1-\rho) \pol, \\ 
			\label{equ:tie_1_rho0}
		J_0 &= \ds(\rho)\p_x \rho_0 + \rho_0 \DD (\rho )\p_x \rho - \pe \rho_0 s(\rho ) \pol,
\end{align}
where $J,J_0$ are integration constants.  Setting $\p_t m =0$ in the equation of motion for $m$ we obtain
\begin{equation}
2m = \px \big\{ \ds( \rho ) \px \pol + \pol \DD (\rho ) \px \rho 
	  - \pe \big[ s(\rho ) \pol^2  + \ds ( \rho ) \rho_a \big] \big \}.
	  \label{equ:tie_1_m}
\end{equation}
In the bulk of either phase, we have $\p_x\rho_\sigma=0$, so from \eqref{equ:tie_1_m},  $m=0$ there. Then evaluating Eqs.~(\ref{equ:tie_1},\ref{equ:tie_1_rho0}) in the bulk shows that $J=0=J_0$.
Using this fact and eliminating $m$ between  Eqs.~(\ref{equ:tie_1},\ref{equ:tie_1_rho0}),
 one obtains using \eqref{equ:DD} that $\partial_x \log [\rho_0/(1-\rho)]=0$ so $\rho_0 = \nu(1-\rho)$ for some constant $\nu$.  Integrating this equation over the whole domain, one obtains the value $\nu$ in \eqref{equ:nu}.
Then, combining Eqs.~(\ref{equ:nu-local},\ref{equ:tie_1},\ref{equ:tie_1_m}), we obtain a condition on $\rho(x)$ alone:
\begin{equation}
	\p_x g(\rho,\p_x\rho,\p_x^2\rho) = 0,
	\label{equ:pxg}
\end{equation}
where 
\begin{equation}
	g(\rho,\p_x\rho,\p_x^2\rho) = g_0(\rho) + \Lambda(\rho) (\p_x \rho)^2 - \kappa(\rho) \p_{x}^2 \rho, 
\end{equation}
with $g_0,\Lambda,\kappa$  given in Eqs.~(\ref{equ:g0}-\ref{equ:lam-kap}).

The method of \cite{masonExactHydrodynamicsOnset2023a, solonGeneralizedThermodynamicsMotilityinduced2018, solonGeneralizedThermodynamicsPhase2018} can now be applied directly to \eqref{equ:pxg}.
Within the bulk of the coexisting phases one has $\p_x\rho=0=\p_x^2\rho$, showing that $g_0(\phi_l)=g_0(\phi_v)$. In addition, \eqref{equ:pxg} can be used to construct an effective free energy $\Phi$, from which $\phi_l,\phi_v$ can be fully determined by a common tangent construction, see Appendix~\ref{sec:coexist} for details.  The compositions of the phases are then given by \eqref{equ:nu-local}.  A numerical implementation of this procedure yields the binodal curves in Fig.~\ref{fig:phase}.

\paragraph{Linear Stability of Homogeneous Solutions} To analyse the stability of homogeneous stationary solutions of \eqref{equ:main}, we introduce a perturbation to $\rho_\sigma$ constant of the form $\delta A^\sigma \exp(\lambda t + i q x t)$ for $\delta \ll 1$, and $\lambda$ can be obtained as the eigenvalue of a $3\times 3$ matrix.  If $\text{Re}(\lambda) > 0$, then the perturbation grows, signaling that the homogeneous state is unstable.
The spinodal is the boundary between the regions of stable and unstable homogeneous solutions.
Full details are given in~Appendix~\ref{sec:linear_stab}.

\paragraph{Traveling solutions via the Method of Matched Asymptotics.}
To analyze T solutions in large domains, we define $\epsilon = 1/L$ and seek a systematic approximation of \eqref{equ:travelling} as $\epsilon\ll 1$ via the method of matched asymptotic expansions. Recall that $z\in [-L/2, L/2]$ and that, without loss of generality, the interface is centered at $z= 0$ (see Fig.~\ref{fig:convergence}).  We assume there is only one interface;  the generalization to multiple interfaces is straightforward. 

We define  the outer region $|z| \gg 1$, where we set $z = \epsilon^{-1} \bar z $ and define $\varrho(z) = \bar \varrho(\bar z)$ and $\poltravel(z) = \bar \poltravel(\bar z)$ so \eqref{equ:travelling} becomes
\begin{multline}\label{equ:outer}
-\epsilon^2 c \bar \varrho_\sigma' =  \epsilon^2 \partial_{\bar z}\left[ \ds( \bar \varrho ) \bar \varrho_\sigma' + \bar \varrho_\sigma \DD (\bar \varrho )  \bar \varrho' \right]  \\
	 - \epsilon \pe \partial_{\bar z} \left[ \bar \varrho_\sigma s( \bar \varrho ) \bar \poltravel  + \sigma \ds (\bar  \varrho ) \bar \varrho_\sigma \right]  - \sigma \bar \poltravel,
\end{multline}
with periodic boundary conditions at $\bar z = \pm 1/2$. 
Expanding $\bar \varrho_\sigma$ and $\bar \poltravel$ in powers of $\epsilon$, $\bar \varrho_\sigma \sim \rhoout{\sigma}{0}(\bar z)  + \epsilon \rhoout{\sigma}{1}(\bar z) + \cdots$ and $\bar \poltravel \sim \bar \poltravel^{(0)} + \epsilon \bar \poltravel^{(1)} + \epsilon^2 \bar \poltravel^{(2)} + \cdots$, we find that the leading- and first-order of \eqref{equ:outer} lead to $\bar \poltravel^{(0)} = 0$ and $\bar \poltravel^{(1)} = - (\pe/2) \partial_{\bar z}[\ds (\rhoout{ }{0}) \rhoout{a}{0} ]$, respectively. The $O(\epsilon^2)$ of \eqref{equ:outer} is
\begin{multline}\label{mae0}
\hspace*{-\multlinegap}- c \bar{\varrho}_\sigma^{(0)\prime} =   \partial_{\bar z}\left[ \ds( \bar \varrho^{(0)} )  \bar \varrho_\sigma^{(0)\prime} + \bar \varrho_\sigma^{(0)} \DD (\bar \varrho^{(0)} )  \bar \varrho^{(0)\prime} \right] - \sigma \bar \poltravel^{(2)}  \\
	 -  \pe \partial_{\bar z} \! \left[ \bar \varrho_\sigma^{(0)} s( \bar \varrho^{(0)} ) \bar \poltravel^{(1)}  \!+\! \sigma \ds (\bar  \varrho^{(0)} ) \bar \varrho_\sigma^{(1)} \! +\! \sigma \ds' (\bar  \varrho^{(0)} ) \bar \varrho^{(1)}  \bar \varrho_\sigma^{(0)}\right],
\end{multline}
using that $\ds (\bar  \varrho ) \bar \varrho_\sigma \sim \ds (\bar  \varrho^{(0)} ) \bar \varrho_\sigma^{(0)} + \epsilon \ds (\bar  \varrho^{(0)} ) \bar \varrho_\sigma^{(1)} + \epsilon \ds' (\bar  \varrho^{(0)} ) \bar \varrho^{(1)}  \bar \varrho_\sigma^{(0)}$ and $\bar \poltravel^{(0)} = 0$. Eliminating $\bar \poltravel^{(2)}$ from \eqref{mae0} by adding them for $\sigma = \pm$ leads to
\begin{multline}\label{mae1}
\hspace*{-\multlinegap}- c  \bar \varrho_a^{(0)\prime} =   \partial_{\bar z}\left[ \ds( \bar \varrho^{(0)} ) \bar \varrho_a^{(0)\prime} + \bar \varrho_a^{(0)} \DD (\bar \varrho^{(0)} )  \bar \varrho^{(0)\prime} \right]  \\
	 -  \pe \partial_{\bar z} \left[ \left( \bar \varrho_a^{(0)} s( \bar \varrho^{(0)} )   +  \ds (\bar  \varrho^{(0)} ) \right) \bar \poltravel^{(1)} \right],
\end{multline}
using again that $\bar \poltravel^{(0)} = 0$. Inserting the expression for $\bar \poltravel^{(1)}$ into Eqs.~(\ref{mae0}, \ref{mae1}) and switching back to $z$ leads to 
\eqref{equ:outer_leading}
at leading order in $\epsilon$.

To solve \eqref{equ:outer_leading}, it remains to determine the boundary conditions as we approach the interface, or $|\bar z|\to 0$. 
In the inner region $z=O(1)$, define densities $\hat\varrho_\sigma(z) =  \varrho_\sigma(z)$. They solve \eqref{equ:travelling}, together with the matching condition to the outer region, $ \lim_{z\to \pm \infty} \hat \varrho(z) = \lim_{\bar z \to 0^\pm} \bar \varrho (\bar z)$; these limits exist under the assumptions of a localized interface.
At leading order in $\epsilon$, the LHS of \eqref{equ:travelling} vanishes: this ensures consistency with the argument above that traveling phase-separated states do not exist, and justifies the scaling of the speed as $c/L$ in \eqref{equ:travelling}. 
Hence, the leading-order inner solution solves the same equations as the PS state [Eqs.~(\ref{equ:tie_1},\ref{equ:tie_1_rho0},\ref{equ:tie_1_m}) with $J=0=J_0$], and interfaces in T states connect points on the binodal curve. 

\paragraph{Numerical Methods} \label{sec:num}
\mbox{}\\ 
\indent $\bullet$ \emph{Particle model}:
The APLG is a continuous time Markov chain on a finite state space. We simulate it exactly with the Gillespie algorithm \cite{erbanStochasticModellingReaction2020}, initially placing a $\sigma$-particle on a lattice site with probability $\phi_\sigma$, where $\phi_{\pm} = \phi_a/2, \phi_0 = \phi_p$.  
In Figs. \ref{fig:MIPS},\ref{fig:CP}, the simulated domain has $\ell_y=\ell_x/4$; we plot the $y-$averaged values of the mesoscopic densities (see \eqref{meso-density}) with radius $r = 0.1$. 

$\bullet$ \emph{Time-stepping of Eq.~\eqref{equ:main}}:
We use a first-order finite-volume scheme in space and forward Euler with adaptive time-stepping in time to obtain time-dependent numerical solutions $\rho_\sigma(x,t)$ to the hydrodynamic PDE \eqref{equ:main}, building on the numerical scheme of \cite{masonExactHydrodynamicsOnset2023a,brunaPhaseSeparationSystems2022} (see Appendix~\ref{sec:timedep}).

$\bullet$ \emph{Classification of the steady-states}: We classify solutions of \eqref{equ:main} into Homogeneous (H), Phase-separated (PS), Traveling (T), or Counter-propagating (CP) using the following two metrics: the approximate speed $\tw{c}(t) := \Vert \pt \rho_\sigma \Vert_2 / \Vert \px \rho_\sigma \Vert_2$ and the distance from uniform 
\begin{equation}
	d_\text{H}(t) := \Big  ( \sum_\sigma
\Vert \rho_\sigma(\cdot,t) - \phi_\sigma \Vert_2 \Big )^{1/2}.
\end{equation}
We solve \eqref{equ:main} until a final time $t^*\ge 700$ when one of the following conditions is satisfied (where $\mathcal T^* = [t^*-500, t^*]$):
\begin{enumerate}[label=(\roman*),itemsep=0pt]
	\item $\sup_{t \in \mathcal T^*} d_\text{H}(t) < 0.05$ $\to$  H solution.
	\item $d_\text{H}(t^*) \geq 0.05$ and $\sup_{t \in \mathcal T^*} \tw{c}(t) < 0.01$ $\to$ PS solution.
	\item $d_\text{H}(t^*) \geq 0.05$, $\tw{c}(t^*) \geq 0.01$ and  $\sup_{t \in \mathcal T^*} \vert  \tw{c}'(t) \vert < 10^{-5}$ $\to$ T solution.
	\item $d_\text{H}(t^*) \geq 0.05$, $\tw{c}(t^*) \geq 0.01$ and  $\sup_{t \in \mathcal T^*} \vert  \tw{c}'(t) \vert \geq 10^{-5}$ $\to$ CP solution.
\end{enumerate}

$\bullet$ \emph{Traveling profiles}:
The profiles $\varrho_\sigma(z)$ and speed $c$ in Fig.~\ref{fig:outer} are obtained numerically by discretizing Eqs.~(\ref{equ:travelling},\ref{equ:outer_leading}) with second-order centered differences and solving the resulting zeroth-finding problem subject to mass constraints $\int \varrho_0 \dd z = \phi_p$ and $\int \varrho_+ \dd z = \int \varrho_- \dd z = \phi_a/2$ in $[-L/2, L/2]$. 
The finite-$L$ system \eqref{equ:travelling} is initialized with a preexisting T solution with similar parameters or a steady state of \eqref{equ:main} and solved subject to periodic boundary conditions and $\varrho(0) = \phi$ without loss of generality (since the problem has translational symmetry).
For $L\gg 1$, the numerical procedure to solve \eqref{equ:outer_leading} is as follows. Given parameter values $\phi$ and $\nu$ \eqref{equ:nu}, we determine the inner solution via the abovementioned Coexisting Phases procedure. This results in liquid $(\varrho_\sigma)_l$ and vapor $(\varrho_\sigma)_v$ values, which become boundary values for the outer problem \eqref{equ:outer_leading}. [If $\phi_v(\nu) = \phi_l(\nu)$, it means that there is no inner region and we may proceed to solve \eqref{equ:outer_leading} subject to periodic boundary conditions as in the finite $L$ case.] Then \eqref{equ:outer_leading} is discretized using second-order finite-differences and solved in $[0,L]$ subject to Dirichlet boundary conditions $\varrho_\sigma(0) = (\varrho_\sigma)_v$ and $\varrho_\sigma(L) = (\varrho_\sigma)_l$ and mass constraints as above. 
If no solution is found, it indicates there may be a second interface. We initialize with a previous single interface T solution with similar parameters. We then place a new interface at $z= \text{argmax}_z \vert \p_z \varrho \vert$.
In both finite $L$ and $L\gg 1$ cases, we solve the resulting systems of equations using the \texttt{NonlinearSolve.jl} package in Julia (see Appendix~\ref{sec:travelling_num}).

\section*{Data availability}
The simulation data generated in this study have been deposited in the Figshare database under accession code \href{https://doi.org/10.25446/oxford.28881329}{10.25446/oxford.28881329} \cite{figshare2025}.

\section*{Code availability}
The code used in this study has been deposited in the GitHub repository under accession code \href{https://doi.org/10.5281/zenodo.15482553}{mbruna/Nonreciprocal$\textunderscore$APLG} \cite{mason2025}.

\section*{Author contributions}
JM, RLJ, and MB designed research, performed research, and wrote the paper.

\section*{Acknowledgements}

We thank Tal Agranov, Martin Burger, Mike Cates, Clement Erignoux, Sarah Loos, and Johannes Zimmer for helpful discussions.  M. Bruna was supported by a Royal Society University Research Fellowship (grant no. URF/R1/180040). J. Mason was supported by the Royal Society Award (RGF/EA/181043) and the Cantab Capital Institute for the Mathematics of Information of the University of Cambridge. For the numerical work, we used the Julia programming language \cite{Julia-2017} and the following packages and tools: DrWatson.jl \cite{Datseris2020}, DifferentialEquations.jl \cite{rackauckas2017differentialequations}, NonlinearSolve.jl \cite{pal2024nonlinearsolve}.

\vfill\eject
\onecolumngrid

\newcommand{\RR}{\mathbb{R}}
\newcommand{\pz}{\p_{ z}}
\newcommand{\y}{{\mathbf y}}{}
\renewcommand{\u}{{\mathbf u}}
\newcommand{\g }{{\bf g}}
\newcommand{\f }{{\bf f}}
\newcommand{\F}{{\mathbf F}}{}

\renewcommand{\div  }{ \nab \cdot} 
\newcommand{\beq}{\begin{equation}}
\newcommand{\eeq}{\end{equation}}

\newcommand{\fp}{F^{\rm p}}  

\newcounter{equationSM}
\newcounter{figureSM}
\newcounter{tableSM}
\stepcounter{equationSM}
\setcounter{equation}{0}
\setcounter{figure}{0}
\setcounter{table}{0}
\setcounter{section}{0}
\makeatletter

\renewcommand{\theequation}{S\arabic{equation}}
\renewcommand{\thefigure}{S\arabic{figure}}
\renewcommand{\thesection}{\Alph{section}} 
\renewcommand{\thesubsection}{\arabic{subsection}} 

\begin{center}
  {\bf\large Appendices}
\end{center}

\newtheorem{definition}{Definition}
\theoremstyle{definition}
\newtheorem{example}{Example}

\section{Hydrodynamic limit} \label{sec:hydro}
The hydrodynamic system of PDEs \eqref{equ:main} describes the local density of each type of particle as the lattice spacing $h \to 0$. For ease of reference, we rewrite \eqref{equ:main} below in terms of $\rho = \rho_+ + \rho_- + \rho_0$, $\rho_a = \rho_+ + \rho_-$ and $\pol = \rho_+-\rho_-$:
\begin{subequations} \label{hydro_SI}
	\begin{align} \label{equ:main_app_1}
	\pt \rho &=  \div \nab \rho
	 - \pe \px \left[ (1-\rho) \pol \right], 
\\ \label{equ:main_app_2}
\pt \rho_a &=  \div \left[ \ds( \rho ) \nab \rho_a + \rho_a \DD (\rho ) \nab \rho \right]
	 - \pe \px \left[ \rho_a s(\rho ) \pol  + \ds ( \rho ) \pol \right],
\\ \label{equ:main_app_3}
\pt \pol &=  \div \left[ \ds( \rho ) \nab \pol + \pol \DD (\rho ) \nab \rho \right]
	 - \pe \px \left[ s(\rho ) \pol^2  + \ds ( \rho ) \rho_a \right]-2\pol.
\end{align}
\end{subequations}

To obtain these equations from the particle-based dynamics of the APLG, we define $\eta_\sigma(\x,t)$ to be $1$ if there is a $\sigma$-particle at position $\x$ and $0$ otherwise. 
The local density is formally  defined as the mean number of particles in a mesoscopic box of radius $r$ around $\x$,
\begin{equation} \label{meso-density}
	\hat\rho_\sigma(\x,t) \approx 
	\frac{1}{(2 r/h + 1)^2} 
	\sum_{\Vert \x - \y \Vert_\infty < r} 
	\eta_\sigma(\y,t),
\end{equation}
for $1 \gg r \gg h$.  
Recalling that the lattice is embedded in a physical domain of size $\ell_x\times \ell_y$ and that the total numbers of active (and passive) particles per site are $\phi_a$ (and $\phi_p$), the relevant limit is $\h\to0$ at fixed $\ell_x,\ell_y,\phi_a,\phi_p$.  The hydrodynamic limit exists 
if that the random variables $\hat\rho_\sigma$  converge (in probability) to deterministic densities $\rho_\sigma$, which are solutions to the hydrodynamic PDE system \eqref{hydro_SI}.

This is proven in the APLG by generalizing the work of Erignoux \cite{erignouxHydrodynamicLimitActive2021}. That work considered a system of pure active particles with continuously varying orientations $(\cos \theta, \sin \theta)$ with $\theta \in [0, 2 \pi)$ instead of only $\theta = \{0, \pi\}$ used here. In both cases, the proof of the convergence $\hat\rho_\sigma\to\rho_\sigma$ is technically challenging because the models are of nongradient type in the sense of \cite{kipnisScalingLimitsInteracting1998}. It is worth emphasizing that this classification is separate from whether the model can be derived as a gradient flow of an equilibrium free energy. Indeed, the hydrodynamic limit of a mixture of two species undergoing a symmetric simple exclusion process is a gradient flow, but the underlying microscopic model is of nongradient type \cite{quastelDiffusionColorSimple1992}. 
Being of nongradient type means instead that the current $j_{\x,\x+h\e_i}$ from  site $\x$ to $\x+h\e_i$, cannot be written as the discrete difference of a local function $g$
\begin{equation}
	j_{\x,\x+h\e_i} \neq  g_{\x+h\e_i}(\eta) - g_{\x}(\eta). 
\end{equation}
The original proof in \cite{erignouxHydrodynamicLimitActive2021} is extremely technical and long; for an abridged version, we refer the reader to \cite[Sec.~5]{masonExactHydrodynamicsOnset2023a}.

The nongradient method \cite{quastelDiffusionColorSimple1992} involves projecting the current onto a space of discrete differences and proving that it can be replaced by its local average in the hydrodynamic limit, e.g., the instantaneous current of $\sigma$-particles going between neighboring sites $\x$ and $\x + h \e_1$ can be approximated as
\begin{equation}\label{approx}
	j_{\x,\x+h{\e}_1}^\sigma \simeq \ds(\rho)[\eta_\sigma(\x+h\e_1)-\eta_\sigma(\x)] + \DD(\rho)[\eta(\x+h\e_1)-\eta(\x)] + \pe \left[\sigma \ds(\rho)\eta_\sigma(\x) + s(\rho) m \eta_\sigma(\x)\right],
\end{equation}
see \cite[\S5(c)]{masonExactHydrodynamicsOnset2023a} for details.
It is the symmetric part of the dynamics (shared by the active and passive particles and leading to the first two terms in the right-hand side of \eqref{approx}) of the APLG that makes the model of nongradient type. As a result, adding a different type of particles (with identical symmetric jump rates) does not bring new challenges. As such, while \cite{erignouxHydrodynamicLimitActive2021, masonExactHydrodynamicsOnset2023a} cannot be used verbatim, the proof of the APLG hydrodynamic limit is a straightforward generalization of those works.

Another signature of the model being of nongradient type is that its hydrodynamic limit contains transport coefficients whose dependence on the density is not known, albeit they are characterized by a variational formula over an infinite-dimensional space \cite{arita2017variational}. This is the case for the self-diffusion coefficient $d_s(\rho)$ in Eq.~\eqref{hydro_SI}. In our analysis, we use the approximation \eqref{eq_ds_approx} of $d_s(\rho)$  obtained in \cite{masonMacroscopicBehaviourTwoSpecies2023}, exploiting a rigorous recursive approach proposed in \cite{landim2001symmetric}. The idea equates to performing a Taylor expansion of $d_s(\rho)$ around $\rho= 0, 1$ up to the first order and combining the two linear approximations in the minimal cubic polynomial. At the level of the self-diffusion coefficient, the mean-field approximation $\langle \eta_\sigma(\x,t)\eta_\sigma(\hat \x,t) \rangle \approx  \rho_\sigma(\x,t)\rho_\sigma(\hat \x,t)$ would result in $d_s(\rho) \equiv 1-\rho$ in \eqref{eq_ds_approx} and, in turn, $\mathcal D(\rho) \equiv 1$ and $s(\rho) \equiv 0$ in \eqref{equ:DD}.

\subsection{Choices of scaling limit} \label{sec:scaling}

As discussed in the main text, the microscopic rates of the APLG exhibit a nontrivial dependence on $h$. For instance, the hopping rate scales as $O(h^{-2})$, while the orientation flip rate is $O(1)$. In the hydrodynamic limit $h \to 0$, the rapid hopping effectively mixes the system, a crucial ingredient in the proof of \eqref{approx}. Consequently, the densities $\rho_+$ and $\rho_-$ emerge as hydrodynamic (slow) fields. This behavior contrasts with NRCH-like frameworks for active-passive mixtures~\cite{youNonreciprocityGenericRoute2020,stenhammarActivityInducedPhaseSeparation2015,wysockiPropagatingInterfacesMixtures2016}, where the hydrodynamic fields would typically be the total active density $\rho_a$ (summed over orientations) and the passive density $\rho_0$, while the magnetization $m$ would be a fast (nonhydrodynamic) field.  

Physically, the magnetization field is ``fast'' if the timescale for orientational relaxation is comparable with the time a particle takes to diffuse a distance comparable with its own size.  The rigorous mathematical analysis of~\cite{masonExactHydrodynamicsOnset2023a} is not possible in this case: it relies on a separation of time scales between hopping and orientational relaxation.  In this sense, the time-scale separation of the APLG is an idealized modeling assumption, which is widespread in interacting particle systems \cite{martin2307transition, agranovThermodynamicallyConsistentFlocking2024, kourbane-housseneExactHydrodynamicDescription2018, erignouxHydrodynamicLimitActive2021}. Nevertheless, the APLG supports complex behavior reminiscent of other active-passive mixtures without any such timescale separation~\cite{agranovThermodynamicallyConsistentFlocking2024,frohoff-hulsmannSuppressionCoarseningEmergence2021,frohoffResonance2023,frohoffUniversal2023}.  That is, the simplified APLG model still captures the emergent behavior of its more complex counterparts.

We also comment on the thermodynamic limit, which is distinct from the hydrodynamic limit.  The thermodynamic limit takes a large system $\ell_x, \ell_y \to \infty$ with all other parameters (including $h$) held fixed. Such scenarios are commonly studied in physics; see~\cite{yao2024} for a recent purely active example that shares strong similarities with the APLG.
The analysis of~\cite{yao2024} does not involve any time scale separation between orientational relaxation and local mixing, which means that fluctuations affect the observed behavior, leading (for example) to bubbly phase separation.  Such effects also lead generically to critical points with nontrivial (non-Gaussian) scaling exponents.  Neither of these effects can be derived from noise-free hydrodynamic equations like \eqref{equ:main}, which describe the APLG. The reason is that the hydrodynamic scaling limit of the APLG is constructed in a way that suppresses fluctuations. This allows for exact analysis but suppresses fluctuation-dominated effects like non-Gaussian critical exponents.

\section{Nonreciprocity} \label{sec:nonreciprocity}
%
This section explains how the hydrodynamic system \eqref{hydro_SI} corresponds to a system with nonreciprocal effective interactions.  We briefly review the origin of the ``nonreciprocal'' terminology in violations of Newton's third law in finite systems of interacting particles.   After that, we discuss nonreciprocal effective interactions for interacting density fields.

\subsection{Ordinary differential equations} \label{sec:nonrec_ODE}

We analyse a finite system of $N$ interacting particles, focusing on the behaviour near a fixed point (see~\cite{fruchartNonRecip2021} for an analysis of bifurcations and phase transitions).
The particles have co-ordinates $x_1,x_2,\dots,x_N$ and follow deterministic overdamped dynamics.  
Write $\textbf{x}=(x_1,x_2,\dots)$ and define an energy $E=E(\textbf{x})$.
We assume that $E$ is translation-invariant, that is 
\begin{equation}
\sum_i \frac{\partial E}{\partial x_i}=0 \; .
\label{equ:trans-inv}
\end{equation} 
This assumption simplifies the analysis by ensuring that all forces can be attributed to interparticle interactions, whose behaviour is consistent with Newton's third law, see below.  (There are no ``body forces'' or external forces.) More general cases can also be analysed in a similar way.

The overdamped dynamics for a reciprocal system is
\beq 
\dot{\textbf{x}} = - \Gamma^{-1} \nabla E,
\label{equ:x-recip}
\eeq
where $\Gamma$ is the friction matrix, which is real, symmetric, and positive definite. 
The energy $E$ is nonincreasing under the dynamics \eqref{equ:x-recip}. 
Assume that $\textbf{x}=0$ is a fixed point [$\nabla E(0)=0$] and linearise \eqref{equ:x-recip}: 
\beq
\dot{\textbf{x}} = B \textbf{x}, \qquad B=-\Gamma^{-1}_0 H_0,
\label{equ:x-linear}
\eeq
where $H_0$ is the Hessian of $E$ at $\textbf{x}= 0$ and $\Gamma_0=\Gamma(0)$.  Stability depends on the eigenvalues of $B$.  Using that $\Gamma_0$ is real symmetric positive definite and similarity-transforming with $\Gamma_0^{1/2}$, the eigenvalues of $B$ are those of the real symmetric matrix $-\Gamma_0^{-1/2} H_0 \Gamma_0^{-1/2}$, hence they are always real.  If the eigenvalues of $H_0$ are positive (minimum of $E$), then the eigenvalues of $B$ are negative (stable fixed point, as expected).  The translation invariance \eqref{equ:trans-inv} means that $H_0$ has a zero eigenvalue, but this may always by removed by defining positions relative to the center of mass.

We have from \eqref{equ:trans-inv} that $\sum_i (H_0)_{ij}=0$ which allows \eqref{equ:x-linear} to be written in components as
\begin{equation}
(\Gamma_0 \dot{\textbf{x}})_i  =  \sum_{j (\neq i)} \fp_{ij},  \qquad 
\fp_{ij} = (H_0)_{ij} (x_i - x_j) \; ,
\label{equ:xfH}
\end{equation}
where we identify $\fp_{ij}$ as the (pairwise) force on particle $i$ from particle $j$.  The Hessian is symmetric, so this recovers Newton's third law as
\begin{equation}
\fp_{ij} = - \fp_{ji}
\label{equ:newton}
\end{equation}
(Note: we did not assume that the energy is a sum of pairwise additive contributions, this formula giving $\fp_{ij}$ in terms of the Hessian is always valid close to the fixed point.)

To generalise \eqref{equ:x-recip} to non-reciprocal systems, we write
\beq \label{generic_ODE}
\dot{\textbf{x}} = \Gamma^{-1} \F,
\eeq
where $\F$ is a force, which is not generically the gradient of any energy $E$. We continue to assume that $\textbf{x}=0$ is a fixed point and translation invariance (now in the form $\sum_j (\partial F_i/\partial x_j)=0$ for all $i$). Linearising about the fixed point gives the analog of \eqref{equ:xfH}
\begin{equation}
(\Gamma_0 \dot{\textbf{x}})_i  =  \sum_{j (\neq i)} \fp_{ij},  \qquad 
\fp_{ij} = \frac{\partial F_i}{\partial x_j} (x_j - x_i) 
\label{equ:xfJ}
\end{equation}
where again $\fp_{ij}$ is the force on particle $i$ from particle $j$.
If ${\bf F}=-\nabla E$ then we recover the previous case, including Newton's third law~\eqref{equ:newton},  
but this is violated if $\frac{ \partial F_i }{ \partial x_j } \neq \frac{ \partial F_j }{ \partial x_i }$.
Striking examples occur when particle $i$ is attracted towards particle $j$, but particle $j$ is repelled from particle $i$, leading to predator-prey-type dynamics \cite[Chapter 3]{murray2007mathematical}.    

It is also useful to write the linearised system \eqref{equ:xfJ} as 
\beq
\dot{\textbf{x}} = B_{\rm NR} \textbf{x}, \qquad B_{\rm NR} = \Gamma_0^{-1} J, \qquad J_{ij} = \frac{\partial F_i}{\partial x_j}
\label{equ:NBR}
\eeq
analogous to \eqref{equ:x-linear}.  The matrix $B_{\rm NR}$ may have complex eigenvalues (because $J$ is not symmetric in general).
We note that for any given equation of motion \eqref{generic_ODE}, it may not be obvious whether it can be factorised as \eqref{equ:x-recip}.  However, if linear stability yields complex eigenvalues, then such a factorisation is not possible, and the system must be nonreciprocal.

\begin{example}[Non-reciprocal ODE system]
Consider three particles with positions $x_1, x_2, x_3$ moving in the interval $[0, 2\pi)$ with periodic boundaries.  We take as the equation of motion
\begin{equation}
\begin{pmatrix} \dot x_1 \\ \dot x_2 \\ \dot x_3 \end{pmatrix} = 
\begin{pmatrix} (1+a) \sin(x_2-x_1) + (1-a) \sin(x_3-x_1)  \\ 
 (1+a) \sin(x_3-x_2) + (1-a) \sin(x_1-x_2) \\ 
 (1+a) \sin(x_1-x_3) + (1-a) \sin(x_2-x_3) \end{pmatrix}
\end{equation}
where the friction matrix is $\Gamma_0=\mathbf{1}$.  The fact that forces only depend on particle separations ensures translation invariance, and allows individual terms to be easily identified as interparticle forces. 

For $a=0$ this is a reciprocal system with $E=-[\cos(x_1-x_2)+\cos(x_1-x_3)+\cos(x_2-x_3)]$: all forces are attractive and setting $x_1=x_2=x_3$ gives a fixed point.  For $0<|a|\leq 1$ some attractive forces are weakened while others are strengthened, which breaks reciprocity; for $|a|>1$ some forces become repulsive so that (for example) particle $1$ may be attracted to particle $2$ while particle $2$ is repelled from particle $1$.  Nevertheless, the fixed point remains the same. 

Linearising about this point we arrive at the situation described in \eqref{equ:xfJ}, with
\begin{equation} 
\fp_{12} = (1+a)(x_2-x_1), \qquad \fp_{21} = (1-a)(x_1-x_2),
\end{equation}
being the forces between particles $1$ and $2$.
Clearly $\fp_{12}\neq -\fp_{21}$ and Newton's third law is violated, unless $a=0$.  The situation is similar for forces between other pairs of particles.  Writing the linearised equation of motion as in \eqref{equ:NBR} gives
\begin{equation}
B_{\rm NR} = 
\begin{pmatrix} 
-2 & 1+a & 1-a \\ 
1-a & -2 & 1+a \\
1+a & 1-a & -2 
\end{pmatrix}
\end{equation}
whose eigenvalues are $0,-3\pm ia\sqrt{3}$.  The fixed point is (marginally) stable in the reciprocal case and this is unchanged by non-reciprocity.  However, any non-zero $a$ yields complex eigenvalues, indicating oscillatory decay to the fixed point.  The existence of  oscillations for all $a\neq0$ is not generic: it occurs because the underlying reciprocal system has degenerate eigenvalues.  (It is straightforward to construct similar systems where the eigenvalues remain real for some finite range of $a$.)  However, complex eigenvalues are not compatible with reciprocal forces.
\end{example}

\subsection{Cross-diffusion systems}

Now consider two species with densities $\rho_1$ and $\rho_2$ that vary in space.  A generic reciprocal PDE system (for example describing relaxation towards thermal equilibrium) analogous to \eqref{equ:x-recip} is:
\beq
\partial_t \begin{pmatrix}  \rho_1 \\  \rho_2 \end{pmatrix} 
 = 
 \nabla \cdot \left[ M \nabla 
 \begin{pmatrix}\delta  E/\delta \rho_1  \\ 
\delta E/\delta \rho_2
  \end{pmatrix}  \right],
\label{equ:rho-recip}
\eeq
where the mobility matrix $M$ is real symmetric positive definite (in general it may depend on $\rho_1,\rho_2$), and $E$ is the free energy. 
Assuming for simplicity that $E = \int \varepsilon(\rho_1,\rho_2) \mathrm{d} x$ so that $\delta  E/\delta \rho_i = \partial\varepsilon/\partial\rho_i$ for $i = 1,2$, \eqref{equ:rho-recip} reduces to the cross-diffusion problem
\beq
\partial_t \begin{pmatrix}  \rho_1 \\  \rho_2 \end{pmatrix} 
 = 
 \nabla \cdot \left[ M H  
 \begin{pmatrix} \nabla \rho_1 \\ \nabla \rho_2
  \end{pmatrix}  \right],
\label{equ:rho-almost-linear}
\eeq
where $H$ is the Hessian matrix of $\varepsilon$.  
The steady states $\rho_i^\infty$ of \eqref{equ:rho-almost-linear} are the minimisers of $E$ and satisfy $\partial \varepsilon/\partial \rho_i$ constant. 
For $M$ linear, this is the set-up of Onsager's reciprocity principle \cite{PhysRev.38.2265}, in which the macroscopic symmetry of a system close to equilibrium arises from its microscopic time reversibility, see also \cite{MielkeRengerPeletier+2016+141+149}.

Linearising \eqref{equ:rho-almost-linear} about $\rho_i^\infty$ and taking $\rho_i = \rho_i^\infty + e^{-iqx} f_i(t)$ gives 
\beq 
\begin{pmatrix} \dot f_A \\ \dot f_B \end{pmatrix} 
 = 
B
 \begin{pmatrix} f_A \\  f_B
  \end{pmatrix} 
  , \qquad B = -q^2 M_\infty H_\infty,
\label{equ:rho-linear}
\eeq
where $M_\infty = M(\rho_1^\infty, \rho_2^\infty)$ and similary for $H_\infty$.  Eq.~\eqref{equ:rho-linear} is analogous to \eqref{equ:x-linear} and, for the same reasons, the eigenvalues of $B$ are real and, if $H_\infty$ is positive definite, negative (indicating a linearly stable fixed point). 

To break the reciprocal structure, we generalise \eqref{equ:rho-almost-linear} [by analogy with \eqref{generic_ODE}] as
\beq
\partial_t\begin{pmatrix}  \rho_1 \\  \rho_2 \end{pmatrix} 
 = 
 \nabla \cdot \left[ {\cal B } 
 \begin{pmatrix} \nabla \rho_1 \\ \nabla \rho_2
  \end{pmatrix}  \right],
  \label{equ:rho-V}
\eeq
where  ${\cal B}={\cal B}(\rho_1,\rho_2)$ is a matrix.
It is not obvious in general whether such an equation can be factorised into the reciprocal form \eqref{equ:rho-almost-linear}.  However, if the eigenvalues of ${\cal B}$ are complex the system can not be factorised in this way. 

Physically, the most common signature of nonreciprocal effective interactions is that species 1 is attracted to species 2, while species 2 is repelled from species 1.  This intuitively corresponds to ${\cal B}_{12}$ and ${\cal B}_{21}$ having opposite signs.  Note however that this difference of signs may also occur in reciprocal systems,\footnote{For example, $M H = \begin{pmatrix} 8 & 3 \\ 3 & 8\end{pmatrix} \begin{pmatrix} 4 & -1 \\ -1 & 2\end{pmatrix}
= \begin{pmatrix} 29 & -2 \\ 4 & 13\end{pmatrix} = {\cal B}$.} so it cannot be regarded as a definite signature of nonreciprocity.

In order to identify which systems are nonreciprocal, we write  \eqref{equ:rho-almost-linear}  as
\beq
\partial_t \begin{pmatrix}  \rho_1 \\  \rho_2 \end{pmatrix} 
 = 
 -\nabla \cdot \left[ M \begin{pmatrix} \F_1 \\  \F_2
  \end{pmatrix}  \right]
, \qquad
\begin{pmatrix} \F_1 \\  \F_2
  \end{pmatrix} = - G  \begin{pmatrix} \nabla \rho_1 \\ \nabla \rho_2
  \end{pmatrix},
\label{equ:rho-a}
\eeq
where $G$ is a nonsymmetric matrix but $M$ is still symmetric positive definite, and $\F_1, \F_2$ are the ``forces'' on species 1, 2:
 If $G_{12}<0$ and $G_{21}>0$ then species 1 feels a force towards 2, while species 2 feels a force away from 1. 
  The systems that we consider reduce to equilibrium (reciprocal) systems in some suitable limit in which $M$ is known.  We suppose the nonreciprocity enters via the forces $G$ so the matrix $M$ is fixed to its equilibrium form.   Then:
\begin{definition}[Nonreciprocal effective interactions in cross-diffusion] Consider a (nonlinear) cross-diffusion system 
\beq \label{general_nonrec}
\partial_t \begin{pmatrix}  \rho_1 \\ \rho_2 \end{pmatrix} 
 = 
 \nabla \cdot \left[ M G \begin{pmatrix}  \nabla \rho_1 \\ \nabla \rho_2
  \end{pmatrix}  \right]
\eeq
where $M$ is a given $2\times 2$ symmetric positive definite matrix.  The system has nonreciprocal effective interactions if $G$ is not symmetric. 
\label{def_nonrec}
\end{definition}
The broken symmetry of $G$ is analogous to taking $J$ non-symmetric in the mechanical case of \eqref{equ:NBR}, in which case Newton's third law \eqref{equ:newton} is violated.
The interesting cases tend to happen when the off-diagonal elements of $G$ have opposite signs or $G$ has complex eigenvalues.  
It may be possible to write the force as a gradient of a selfish energy $E^{(i)}$ of species $i$ \cite{avni2023non}, for example $G_{ij} \nabla \rho_j = \nabla(\delta E^{(i)}/\delta \rho_j)$. Note that the definition \eqref{general_nonrec} does not rule out an alternative factorisation of the equation of motion as \eqref{equ:rho-almost-linear} (with some different mobility $M$), see also~\cite{degond1997symm}.  However, if $MG$ has complex eigenvalues, such a factorisation is not possible.

\begin{example}[Nonreciprocal Cahn--Hilliard model \cite{wittkowskiNonequilibriumDynamicsMixtures2017,Dinelli2023,youNonreciprocityGenericRoute2020}]
	In the notation of \eqref{general_nonrec}, Eq. (1) of \cite{youNonreciprocityGenericRoute2020} has $M= \bm{1}$ and 
\begin{equation}
G =  H + \begin{pmatrix} 0 & \kappa-\delta \\ \kappa+\delta & 0 \end{pmatrix},
\end{equation}
where $H$ is the Hessian of $\varepsilon_{\rm CH}$, defined such that $E = \int \varepsilon_{\rm CH}(\rho_1,\rho_2) \dd x$ is the Cahn--Hilliard energy and
\beq
 \varepsilon_{\rm CH}(\rho_1,\rho_2) = \sum_{i=1,2}  \left( \frac{\chi_i}{2} \rho_i^2 + \frac{1}{12} \rho_i^4 + \frac{\gamma_i}{2} |\nabla\rho_i|^2 \right ).
\eeq
The equilibrium baseline for this model is $\delta=0=\kappa$.  For $\delta \ne 0$, the matrix $G$ is not symmetric and the model is nonreciprocal according to Definition~\ref{def_nonrec}.  
Further, for $|\delta| > |\kappa|$, the off-diagonal elements of $G$ have opposite signs and, e.g., species 1 likes species 2 but species 2 dislikes species 1. When $\delta^2 > \kappa^2 + \alpha$, where $\alpha$ is a positive constant that depends on the self-interaction terms in $E$, the eigenvalues of $G$ are complex conjugates and traveling patterns emerge~\cite{youNonreciprocityGenericRoute2020}. 
\end{example}

\begin{example}[APLG model \eqref{hydro_SI}] \label{example3}
The APLG in \eqref{hydro_SI} is not a cross-diffusion system because the densities $\rho_\pm$ are not individually conserved.  To identify the effective nonreciprocal interactions, we consider solutions with macroscopically smooth densities, which obey the outer equations of the matched asymptotic analysis.  Specifically, writing \eqref{equ:outer_leading} in terms of the two conserved densities $\rho_a(x,t), \rho_0(x,t)$ in the fixed reference frame, one obtains a cross-diffusion system
	\begin{align}
\begin{aligned}
	\partial_t \rho_a & = \partial_x\left[ (d_s + \rho_a \mathcal D) \partial_x \rho_a + \rho_a \mathcal D \partial_x\rho_0 + \frac{1}{2} {\rm Pe}^2  (s \rho_a+ d_s) \partial_x (d_s \rho_a)  \right],
\\ 
\partial_t \rho_0 & =  \partial_x\left[ (d_s + \rho_0 \mathcal D) \partial_x \rho_0 + \rho_0 \mathcal D \partial_x \rho_a + \frac{1}{2} \pe^2 s \rho_0 \partial_x (d_s \rho_a) \right],
\end{aligned}
\label{equ:aplg-out}
\end{align}

The corresponding equilibrium model is ${\rm Pe}=0$, where \eqref{equ:aplg-out} is reciprocal with the mobility matrix \cite{quastelDiffusionColorSimple1992, masonMacroscopicBehaviourTwoSpecies2023}
\beq \label{M_pe0}
M = \frac{1-\rho}{\rho} \begin{pmatrix} \rho_a^2 & \rho_a \rho_0 \\ \rho_a \rho_0 & \rho_0^2
\end{pmatrix}  + \frac{\rho_a \rho_0}{\rho}d_s(\rho) \begin{pmatrix}
	1 & -1\\-1 & 1
\end{pmatrix},
\eeq
and $E=\int \varepsilon(\rho_0,\rho_a) dx$, where 
\beq \label{entropyAPLG}
\varepsilon =  \rho_0 \log \rho_0  + \rho_a \log \rho_a  +(1-\rho) \log (1-\rho) .
\eeq
Writing $H$ for the Hessian of $\varepsilon$, we obtain for ${\rm Pe}=0$ 
\beq \label{eq:MH_out}
\partial_t \begin{pmatrix}  \rho_a \\  \rho_0 \end{pmatrix} 
 = 
 \partial_x  \left[ M H  
 \begin{pmatrix} \partial_x \rho_a \\ \partial_x \rho_0
  \end{pmatrix}  \right], \qquad
MH = \frac{1}{\rho} \begin{pmatrix} \rho_a + d_s \rho_0 & \rho_a(1-d_s) \\ \rho_0(1-d_s) & \rho_0 + d_s \rho_a) \end{pmatrix}.
\eeq
Note that, in this case, $MH$ is not symmetric but the system is reciprocal (since $H$ is symmetric).

Then  \eqref{equ:aplg-out} becomes for ${\rm Pe}>0$:
\beq
\partial_t \begin{pmatrix}  \rho_a \\  \rho_0 \end{pmatrix} 
 = 
 \partial_x  \left[ M ( H + \tfrac{\pe^2}{2} G_{\rm NR} )  
 \begin{pmatrix} \partial_x \rho_a \\ \partial_x \rho_0
  \end{pmatrix}  \right], \qquad  M G_{\rm NR}   = \begin{pmatrix}
	(s \rho_a + d_s)(d_s + d_s' \rho_a) & (s \rho_a + d_s) d_s' \rho_a\\
	s \rho_0 (d_s + d_s' \rho_a) &  s \rho_0 d_s' \rho_a
\end{pmatrix},
  \label{equ:mf-M-full}
\eeq
with $M$ still given in \eqref{M_pe0}. Inverting $M$, we find that $G_{\rm NR}$ is not symmetric, signalling that \eqref{equ:aplg-out} is nonreciprocal  for $\pe >0$, according to Definition \ref{def_nonrec}.
\end{example}

\begin{figure}[t]
\centering
\includegraphics[width=.3\linewidth]{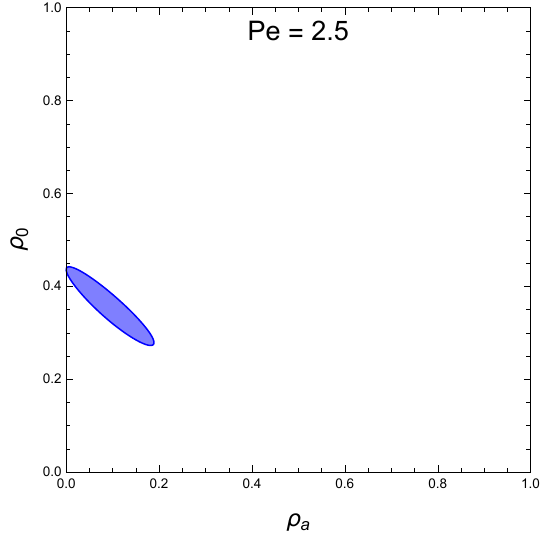}
\includegraphics[width=.3\linewidth]{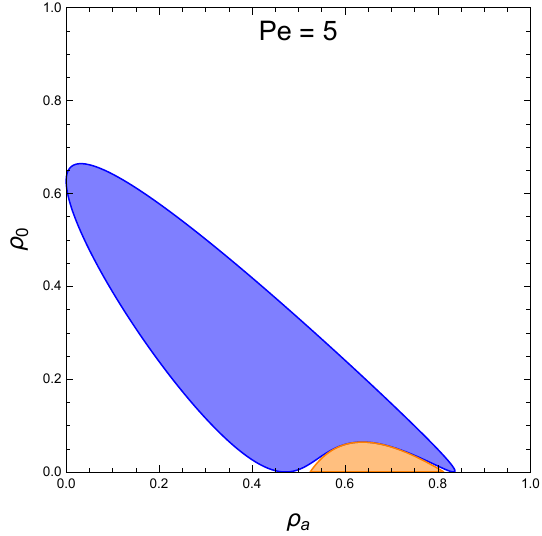}
\includegraphics[width=.3\linewidth]{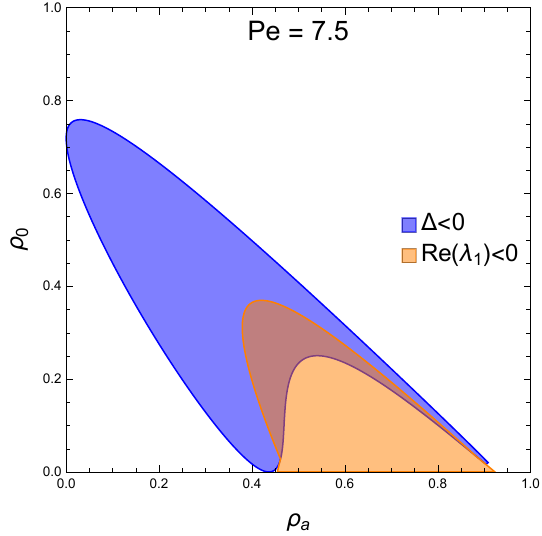}
\caption{\textbf{Spectrum of the cross-diffusion APLG system.}
Illustrative features of the spectrum of $M G = MH + \frac12\pe^2 MG_\text{NR}$ as defined in \eqref{eq:MH_out}, \eqref{equ:mf-M-full}, for various values of $\pe$ ($\pe = 2.5, 5, 7.5$). Denoting the eigenvalues of $MG$ by $\lambda_1, \lambda_2$, the blue region indicates complex eigenvalues (with discriminant $\Delta = (\lambda_1 + \lambda_2)^2 - 4 \lambda_1 \lambda_2 <0$) and the orange region indicates that at least one eigenvalue has negative real part (Re$(\lambda_i)<0$). Where the blue and orange regions overlap, the instability involves traveling and growing sinusoids; otherwise, the orange region corresponds to sinusoids that grow but do not travel.
}
\label{fig:nonrec}
\end{figure}

Pursuing further this example by considering the eigenvalues of the matrix $MG = M( H + \frac12{\rm Pe}^2 G_{\rm NR})$, we find that the spectrum is complex in general (see Fig.~\ref{fig:nonrec}).  For the region with complex eigenvalues, there is no possible factorisation of \eqref{equ:aplg-out} in the manifestly reciprocal form \eqref{equ:rho-almost-linear}.  In practice, complex eigenvalues appear for $\pe \gtrapprox 2.5$ and complex eigenvalues with negative real part (indicating oscillatory instabilities) for $\pe \gtrapprox 5$. 
Note the matrix $MG$ controls the linear stability of homogeneous states but it differs from the matrix of linear stability by a sign, c.f. \eqref{equ:rho-linear}.
Hence the region where at least one eigenvalue of $MG$ has negative real part in Fig.~\ref{fig:nonrec} approximates well the area under the spinodal curves in Fig.~\ref{fig:phases}, see further details in Sec.~\ref{sec:spinodal} below. 

As a final note, we consider the outer problem \eqref{equ:aplg-out} in a mean-field approximation \cite{masonMacroscopicBehaviourTwoSpecies2023}, such that $d_s=1-\rho$ and hence $\mathcal D(\rho) \equiv 1, s(\rho)\equiv 0$.
The system still takes  the form \eqref{equ:mf-M-full} with $H$ as above, and
\beq
M = (1-\rho_a - \rho_0) \begin{pmatrix} \rho_a & 0 \\ 0 & \rho_0
  \end{pmatrix}, 
\qquad 
G_{\rm NR} = \begin{pmatrix} 
-2 + \frac{1-\rho_0}{\rho_a}  & - 1
  \\ 0 & 0  \end{pmatrix},
  \label{equ:mfa_terms}
\eeq
Recalling \eqref{equ:rho-a}, the Pe-dependent force on the active particles is
\beq \label{Fa_Pe}
F_a = \frac{\pe^2}{2} \left( \partial_x\rho_0 + \left[ 2-\frac{1-\rho_0}{\rho_a}  \right] \partial_x\rho_a \right)
\eeq
while there is no such force on the passive particles.
The first term (cross-diffusion) means that active particles feel a force towards passive particles.  In a reciprocal system, the passive particles would feel a corresponding attraction towards the active ones, but this is absent here, so the effective interaction is nonreciprocal.
(The second term in \eqref{Fa_Pe} means that active particles also attract each other if $\rho_a$ is large enough: this is the effective attraction that drives MIPS.)

\section{The method of coexisting phases and the binodal curve} \label{sec:coexist}

Equation (\ref{equ:pxg}) of the main text is $\p_x g=0$ with
\begin{equation}
	g(\rho) = g_0(\rho) + \Lambda(\rho) (\p_x \rho)^2 - \kappa(\rho) \p_{x}^2 \rho, 
	\label{equ:def-gg}
\end{equation}
where
\begin{equation}\label{equ:g0}
	g_0(\rho)= - \pe \Big[(1+\nu)\rho - \nu \Big] \ds(\rho)-  \frac{2}{\pe} \log(1-\rho),
\end{equation}
and 
\begin{equation}
\Lambda(\rho) = \frac{-2 \ds(\rho)}{\pe(1-\rho)^2}, \qquad \kappa(\rho) =  \frac{\ds(\rho)}{\pe(1-\rho)} .
\label{equ:lam-kap}
\end{equation}

We now use the method of Refs.~\cite{masonExactHydrodynamicsOnset2023a, solonGeneralizedThermodynamicsMotilityinduced2018, solonGeneralizedThermodynamicsPhase2018} to derive the densities $(\phi_v, \phi_l)$ of the coexisting phases.
The function $g$ is constant in space; we denote its value by $\bar g$.
As stated in \methodsref, gradients of $\rho$ vanish within the bulk of the coexisting phases  so
\begin{equation}
	g_0(\phi_v) = g_0(\phi_l) = \bar g.
	\label{equ:ggg}
\end{equation}
Next we outline the derivation of the effective free energy $\Phi$ from which $\phi_v,\phi_l$ follow by the common tangent construction, see Ref.~\cite{masonExactHydrodynamicsOnset2023a, solonGeneralizedThermodynamicsMotilityinduced2018, solonGeneralizedThermodynamicsPhase2018} for details.
First define a (one-to-one) function $R(\rho)$ such that $\kappa R'' = -(2 \Lambda + \kappa')  R'$ where primes denote derivatives.  Then, the effective free energy is a function $\Phi(R)$ defined (up to an additive constant) by $\Phi'(R(\rho)) = g_0(\rho)$.  The definition of $R$ is chosen such that
\begin{align}
\p_x [ \kappa(\rho) R'(\rho) (\p_x \rho)^2 ] & =  \kappa'(\rho) R'(\rho) (\p_x\rho)^3 + \kappa(\rho) R''(\rho) (\p_x\rho)^3 + 2\kappa(\rho) ( \p_x^2 \rho )  R'(\rho) \p_x\rho
\nonumber \\
& =  2[  \kappa(\rho) ( \p_x^2 \rho ) - \Lambda(\rho) (\p_x\rho)^2 ] R'(\rho) \p_x \rho,
\label{equ:kappa-magic}
\end{align}
which will be useful below.

To see the common tangent, consider the difference in $\Phi(R(\rho))$ between two points $x_v,x_l$, one in the bulk of each phase.  The density varies in space between the two points as $\rho=\rho(x)$ and we have
\begin{align} \label{equ:Phi-Phi}
	\begin{aligned}
		\Phi(R(\phi_l)) - \Phi(R(\phi_v))  &= \int_{x_v}^{x_l} \Phi'(R(\rho)) R'(\rho) \p_x\rho \, dx
 = \int_{x_v}^{x_l}   g_0(\rho) R'(\rho) \p_x\rho \, dx \\
 &= \int_{x_v}^{x_l} g(\rho,\p_x\rho,\p_x^2\rho) R'(\rho) \p_x\rho \, dx + \int_{x_v}^{x_l}  [ \kappa(\rho) \p_x^2\rho - \Lambda(\rho) (\p_x \rho)^2  ] R'(\rho) \p_x\rho \, \dd x,
	\end{aligned}
\end{align}
where the first equality is the chain rule, the second is the definition of $\Phi$, and the third is \eqref{equ:def-gg}.  
Using \eqref{equ:kappa-magic}, the last integrand in \eqref{equ:Phi-Phi} is a total derivative, that is
\begin{equation}
 \int_{x_v}^{x_l}  [ \kappa(\rho) \p_x^2\rho - \Lambda(\rho) (\p_x \rho)^2  ] R'(\rho) \p_x\rho \, \dd x = \frac12 \left[ \kappa(\rho)R'(\rho) \p_x \rho \right]_{x=x_v}^{x_l} = 0.
\end{equation}
The last equality holds because $\p_x\rho=0$ in the bulk of the phases.  Using this in \eqref{equ:Phi-Phi} and observing that $g=\bar g$ is constant in space, we find
\begin{align}
\Phi(R(\phi_l)) - \Phi(R(\phi_v))  & = \bar g \int_{x_v}^{x_l}   
 \p_x R(\rho) \dd x 
= \bar g [ R(\phi_l) - R(\phi_v) ].
\label{equ:almost-tangent}
\end{align}

Finally, using \eqref{equ:ggg} and the definition of $\Phi$ we have that $\bar g = \Phi'(R(\phi_l)) = \Phi'(R(\phi_v))$ so introducing the shorthand notation $R_l=R(\phi_l)$ and $R_v=R(\phi_v)$, 
\eqref{equ:ggg} becomes  $\bar g = \Phi'(R_l) = \Phi'(R_v)$
and \eqref{equ:almost-tangent} yields
\begin{equation}
\Phi(R_l)  - R_l \Phi'(R_l) = \Phi(R_v)  - R_v \Phi'(R_v),
\label{equ:tangent-P}
\end{equation}
while \eqref{equ:ggg} is
\begin{equation}
\Phi'(R_l) = \Phi'(R_v).
\label{equ:tangent-mu}
\end{equation}
Eqs.~(\ref{equ:tangent-P})-(\ref{equ:tangent-mu}) are exactly the common tangent construction (convex hull) of $\Phi$, as required.  The functions $\Phi$ and $R$ are easily determined from their definitions via numerical integration, so it only remains to solve the two simultaneous equations (\ref{equ:tangent-P})-(\ref{equ:tangent-mu}).  Note in particular that the definition of $R$ can be used together with \eqref{equ:lam-kap} to obtain
\begin{equation}
	R'(\rho) = \frac{1}{\ds(\rho)(1-\rho)^3}.
\end{equation}
(The definition fixes $R'$ up to an arbitrary multiplicative constant, set to unity here.)

\section{Linear stability of homogeneous solutions and the spinodal curve} \label{sec:linear_stab}

The homogeneous solution  $(\rho,\rho_a,m)=(\phi,\phi_a,0)$ is always a solution of Eqs.~\eqref{hydro_SI}.  We analyze the linear stability of this solution by taking $\rho = \phi + \delta \tilde \rho$ , $ \rho_a = \phi_ a + \delta \tilde \rho_a $ , $ m = 0 + \delta \tilde m$ with $y$-independent perturbations $\tilde \rho, \tilde \rho_a, \tilde m$.   At first order in $\delta $, we obtain 
\begin{align}
\label{act_pas_eq_4}
		 \pt \tw{\rho} &=  \p_x^2 \tw{\rho}  -  \text{Pe} \px \left[(1-\phi) \tw{\pol}  \right]   , \\
		\pt \tw{\rho}_a &=  \px  \left[ \ds( \phi ) \px \tw{\rho}_a + \phi_a \DD (\phi ) \px \tw{\rho} \right] -  \text{Pe} \px \left[ \phi_a s(\phi ) \tw{\pol}  + \ds ( \phi ) \tw{\pol} \right],  \\
		\pt \tw{\pol} &=  \px \left[ \ds( \phi ) \px \tw{\pol} \right] 
		  -  \text{Pe} \px \left[ \ds ( \phi ) \tw{\rho}_a + \ds' ( \phi ) \phi_a \tw{\rho} \right]  -2 \tw{\pol}. 
\end{align}
Taking a solution of the form 
\begin{align}
	(\tw{\rho},\tw{\rho}_a,\tw{\pol}) = (A_1, A_2, A_3) \exp(\lambda t + iq x )
	\label{equ:trial-linear}
\end{align}
yields
\begin{align}
\label{equ:sin_pert}
\begin{aligned}
			 \lambda A_1 &= -q^2 A_1 - i q \text{Pe} (1-\phi) A_3 , \\
		\lambda A_2 &= - q^2 \phi_a \DD (\phi ) A_1 -q^2 \ds( \phi )  A_2  - i q \text{Pe}  [ \phi_a s(\phi )   + \ds ( \phi ) ] A_3,  \\
		\lambda A_3 &=  	- i q \text{Pe} \ds' ( \phi ) \phi_a A_1	  - i q \text{Pe} \ds ( \phi ) A_2 -  [q^2 \ds( \phi ) + 2] A_3.
		\end{aligned}
\end{align}
Therefore $\lambda$ is an eigenvalue of the $3\times 3$ matrix
\begin{equation}\label{equ:W_matrix}
	 W = - \begin{pmatrix}
		q^2 & 0 &  i q \pe (1-\phi) \\
		 q^2 \phi_a \DD (\phi ) & q^2 \ds( \phi ) &  i q  \pe  [ \phi_a s(\phi )   + \ds ( \phi ) ] \\
		 i q \pe \ds' ( \phi ) \phi_a &  i q  \pe \ds ( \phi ) & q^2 \ds( \phi ) + 2
	\end{pmatrix}.
\end{equation}
This matrix is not Hermitian, so its eigenvalues are, in general, complex. However, if the spectrum is complex, then two of the eigenvalues form a complex conjugate pair and the other remains real; this is due to symmetry under $\lambda \mapsto \bar \lambda$ and $(A_1, A_2, A_3) \mapsto (\bar A_1, \bar A_2, - \bar A_3)$ in \eqref{equ:sin_pert}, where $\bar A$ denotes the complex conjugate of $A$.

As usual, the homogeneous state is linearly stable if all eigenvalues $\lambda$ have negative real parts.  (It is implicit that $\lambda$ depends on $q$; this condition must hold for all $q$.)  
The resulting instabilities can have several types; we classify them according to the scheme of Ref.~\cite{frohoff-hulsmannNonreciprocalCahnHilliardModel2023}.  Our system has two conserved densities $\rho,\rho_a$ and a nonconserved magnetization $m$.  The behavior is controlled by the conserved densities, which restricts the behavior to four of the eight types considered in Ref.~\cite{frohoff-hulsmannNonreciprocalCahnHilliardModel2023}.  If the dominant eigenvalue of $W$ is real then it is called stationary, else it is called oscillatory; if the instability is initiated by modes with $q\to0$ then it is called large scale, else it is called small scale.  The resulting types are then conserved-Turing (stationary, small scale), Cahn--Hilliard (stationary, large scale), conserved-Hopf (oscillatory, large scale), or conserved-wave (oscillatory, small scale). 

\begin{figure}[tbh]
\centering
\includegraphics[width=.8\linewidth]{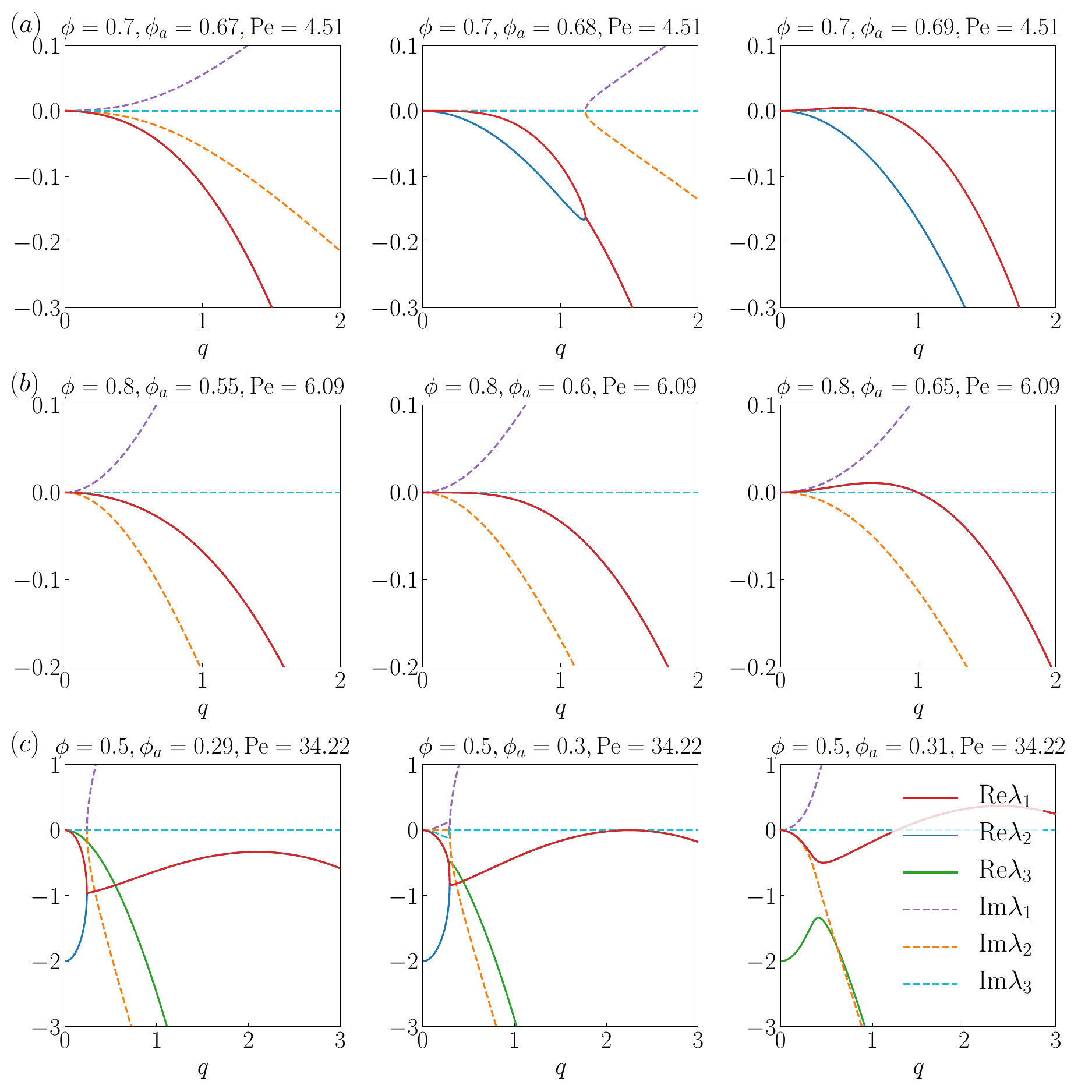}
\caption{\textbf{Linear stability of the APLG system. }
Eigenvalues of $W$ \eqref{equ:W_matrix} at the onset of instability. Active volume fraction $\phi_a$ increases from left to right. The critical volume fraction is displayed in the central column. (a) Cahn-Hiliard instability (stationary, large-scale). (b) Conserved-Hopf instability (oscillatory, large-scale). (c) Conserved-wave instability (oscillatory, small-scale). 
}
\label{fig:eigenvalues}
\end{figure}

Fig.~\ref{fig:eigenvalues} illustrates the range of possible behavior, showing several instabilities that occur on increasing $\phi_a$ at fixed total density $\phi$.  Row (a) shows the onset of a Cahn-Hilliard instability (stationary, large-scale), row  
(b) shows the onset of a conserved-Hopf instability (oscillatory, large-scale), and row (c) shows the onset of a conserved-wave instability (oscillatory, small-scale).  
In the left (right) column, the active volume fraction is sub(super)-critical $\phi_a < \phi_a^*$ where $\max_q \text{Re}\lambda < 0$ ($\phi_a > \phi_a^*$ where $\max_q \text{Re}\lambda > 0$). The central column displays the critical active volume fraction $\phi_a = \phi_a^*$ corresponding to $\max_q \text{Re} \lambda/q^2 = 0$. We show below that these are the only possible scenarios because the onset of a stationary instability must occur on a large scale.

\subsection{Spinodal curves} \label{sec:spinodal}

The main aim of this analysis is to compute spinodal curves, as shown in Fig.~\ref{fig:phase}.  To this end, note that any eigenvalue of $W$ obeys the cubic equation
\begin{equation}
	\lambda^3 - \text{Tr} (W) \lambda^2 + F(W) \lambda - \text{det}(W) = 0,
\end{equation}
where
\begin{equation}
	F(W) = q^2 \big \{ 2  + 2 \ds(\phi) (1+q^2)   + \ds^2(\phi)  (\pe^2 + q^2)  + \ds'(\phi)  \phi_a (1 - \phi)\pe^2  + \ds(\phi) \phi_a  \pe^2 [ \DD(\phi)  -1] \big \}.
\end{equation}
At the boundary of linear stability, then $\text{Re}(\lambda)=0$ for at least one eigenvalue.  There are two situations where this can happen
\begin{enumerate}
	\item[(i)] $\text{det}(W)=0$, corresponding to a vanishing eigenvalue $\lambda=0$ (stationary instability). 
	\item[(ii)] the characteristic polynomial is of the form  $(\lambda^2+F(W))(\lambda-\text{Tr}(W))=0$, with $F(W)>0$ so that $\lambda=\pm i\sqrt{F(W)}$ is pure imaginary.   This situation holds if and only if $F(W)\text{Tr} (W)=\text{det}(W)<0$ (we have always $\text{Tr} (W)<0$ so the final inequality ensures $F(W)>0$). 
\end{enumerate}

In the stationary case (i) one may solve $\det W=0$ for $\phi_a$ to obtain
\begin{equation}\label{equ:lin_stab_re}
	\phi_a^{{\rm st}}(q) =   \frac
	{2 + \ds(\phi)  \text{Pe}^2 + \ds(\phi) q^2}
	{\text{Pe}^2 [\ds(\phi) -(1- \phi)\ds'(\phi) ] } .
\end{equation}
In the oscillatory case (ii), one may similarly solve $\text{det}(W)=F(W)\text{Tr} (W)$ to obtain
\begin{equation} \label{equ:lin_stab_comp}
\phi_a^{{\rm osc}}(q) = \frac{2 \phi \left(1+ d_s(\phi) q^2 \right) \left[2 + 2 d_s(\phi) + d_s^2(\phi) \pe^2 + \left( d_s(\phi) + 1\right)^2 q^2\right]}{\pe^2 \left\{
2 \ds^3(\phi) q^2 + d_s'(\phi) (\phi - 1) \phi (2 + q^2) + \ds^2(\phi) [2 + (\phi -1 ) q^2] +  \ds(\phi) (\phi -1 ) [2 + (1 + d_s'(\phi) \phi) q^2] \right\}}.
\end{equation}

For $\phi_a\to0$, one sees that all eigenvalues of \eqref{equ:W_matrix} have negative real parts: the system is stable and $\rm{det}(W)<0$.  On increasing $\phi_a$ at fixed $\phi$, it follows that the system first becomes unstable at $\phi_a=\phi_a^*$ with
\begin{equation}
\phi_a^* = \inf_q \big[ \min(\phi_a^{{\rm osc}}(q),\phi_a^{{\rm st}}(q)) \big].
\label{equ:spin}
\end{equation}
[For a finite system, the trial solution \eqref{equ:trial-linear} is restricted to $q=2n\pi/L$ with $n\in\mathbb{Z}$ and we should minimize over this discrete set, so $\phi_a^*$ depends in general on $L$.  We consider the limit $L\to\infty$ here, so we take an infimum over $q>0$.]  
Note also that, if the infimum in \eqref{equ:spin} is achieved by $\phi_a^{{\rm osc}}(q)$, then it is certain that $\det(W)<0$ at this point, as required for an oscillatory instability: this holds because $\det(W)$ only changes sign at $\phi_a=\phi_a^{\rm st}(q)>\phi_a^{\rm osc}(q)$ and $\det(W)<0$ for $\phi_a\to0$.
We also observe from \eqref{equ:lin_stab_re} that $\inf_q \phi_a^{{\rm st}}(q) = \phi_a^{{\rm st}}(0)$, which means that if this instability is of stationary type then it is always large-scale, as already asserted above.  On the other hand, the infimum of $\phi_a^{{\rm osc}}(q)$ may occur as $q\to0$ (large-scale oscillatory instability) or at finite $q$ (small-scale oscillatory instability).

Having determined $\phi_a^*$ in this way (and keeping fixed $\phi$),
 the system always remains unstable for all $\phi_a > \phi_a^*$ (because $\det{W}>0$ and $\text{Tr}({W})<0$).  This means that exchanging passive for active particles cannot restore stability, which may be expected on physical grounds.
Hence, the spinodal curve in the $\phi,\phi_a$ plane is given by $\phi_a=\phi_a^*$.

\begin{figure}[tb]
\centering
\includegraphics[width=\linewidth]{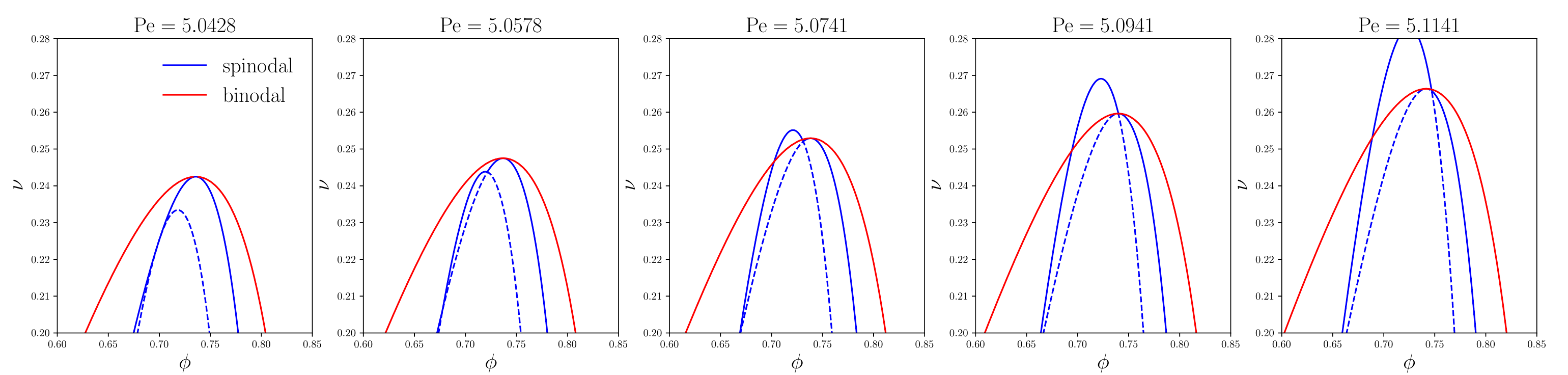}
\caption{\textbf{Protrusion of the spinodal beyond the binodal. } 
Phase diagrams spanned by $\phi$ and $\nu$.
 The spinodal (solid blue) encloses the region of linear stability of homogeneous solutions. 
 The dashed (solid) blue line indicates the minimum (maximum) of \eqref{equ:lin_stab_re} and \eqref{equ:lin_stab_comp}. 
 The binodal (red) encloses the region of phase separation.
 (a-b) spinodal is contained within the binodal  (c-f) spinodal protrudes through the binodal, creating instability in PS solutions. 
}
\label{fig:phases}
\end{figure}

For numerical calculations, it is convenient to parameterize the dependence on $\phi_a$ in terms of the quantity $\nu = \phi_p /(1-\phi)$ defined in Eq.~\eqref{equ:nu}. Then $\nu=\frac{\phi}{1-\phi}$ corresponds to a system of purely passive particles and $\nu=0$ to purely active particles.
Fig.~\ref{fig:phases} illustrates the spinodal in the $(\phi,\nu)$-plane.  We consider a narrow range of $\pe$ in which both oscillatory and stationary instabilities occur, and we plot separately the curves corresponding to $\phi_a=\inf_q\phi_a^{\rm osc}(q)$ and $\phi_a=\inf_q\phi_a^{\rm st}(q)$.  The boundary of linear instability is given by the smaller of these $\phi_a$'s, which corresponds to the larger of the corresponding $\nu$'s.  
The change in the shape of the resulting spinodal curves illustrates the transition between the two types of phase diagrams shown in Fig.~\ref{fig:phases}, as the spinodal curve starts to protrude through the binodal.
[Note that, in regions where $\phi_a^{\rm st}<\phi_a^{\rm osc}$, the dashed curve $\phi_a=\phi_a^{\rm osc}$ does not indicate an eigenvalue with vanishing real part because it lies in a region where $\det(W)>0$.  However, the solid blue (spinodal) curve does always indicate such an eigenvalue.]

\subsection{Protrusion of the spinodal through the binodal}  \label{sec:protrusion}

Since the binodal curve is defined by the common tangent construction on $\Phi$ of Eqs.~(\ref{equ:tangent-P},\ref{equ:tangent-mu}),  
and the function $g_0(\rho)=\Phi'(R(\rho))$ in \eqref{equ:g0}, one sees that $g_0$ must have two turning points between $\phi_l$ and $\phi_v$. 
The equality \eqref{equ:lin_stab_re} that appears in the linear stability analysis is equivalent to $g_0' = 0$.  If the spinodal instability is of stationary type, this means that points of inflection of $\Phi$ correspond to spinodal instabilities, as happens in equilibrium.  
The analog of the critical point in equilibrium occurs when the two turning points coalesce so that $g''=g'=0$ (stationary point of inflection).  At this point, the binodal and spinodal curves are tangent to each other, which is again analogous to equilibrium.  It corresponds to a supercritical pitchfork bifurcation.  The left panel of Fig.~\ref{fig:phases} illustrates this case. 

However, if the instability of the homogeneous state is of oscillatory type, the spinodal is $\phi_a=\phi_a^{\rm osc}$.  This condition has no direct connection with the function $g_0$, so the spinodal is not determined by $\Phi$.  On increasing $\pe$ in Fig.~\ref{fig:phases}, the spinodal protrudes through the binodal for a range of densities, below the critical point.   Further increasing $\pe$, this range extends to cover the critical point itself.
 
This protrusion has two effects. Firstly, where the spinodal protrudes through the low-density (vapor) branch of the binodal, the corresponding PS states also become unstable (because the large domain of the vapor phase behaves like a homogeneous state at the same density).   
Therefore, the steady state must be dynamic as both H and PS solutions are unstable. 
Secondly, when the spinodal engulfs the critical point (the maximum of the binodal in Fig.~\ref{fig:phases}), the H solution becomes unstable on both sides of the bifurcation. At this point, the critical bifurcation changes from supercritical to subcritical. 

As a final comment in this section, note that binodal is analytic, and has a quadratic Taylor expansion near its maximum (in the $\nu,\phi$ plane).  This maximum is a critical point, and the analytic structure is relevant for the critical behaviour of the model, whose exponents are those of the Gaussian fixed point (of the renormalisation group).
This effect can be traced back to the $h$-dependence of the APLG rates, which acts to suppress fluctuations, and enables the exact derivation of the hydrodynamic limit.

\subsection{Small $q$ approximation of the spinodal} \label{sec:approx_spi}
Consider \eqref{equ:sin_pert} in the $L\gg 1$ limit, corresponding to wavenumber $q \ll 1$. From the first two equations in \eqref{equ:sin_pert} we find $\lambda = O(q^2)$, $A_1, A_2 = O(1)$ and $A_3 = O(q)$. As a result, we can solve for $A_3$ in the third equation to obtain
\begin{equation} \label{A3_isolated}
	A_3 =  -\frac{1}{2}  i q \pe \left[ \ds' ( \phi ) \phi_a A_1 + \ds ( \phi ) A_2 \right] + O(q^2).
\end{equation}
The fact that $A_3$ (corresponding to the perturbation $\tilde m$) drops for $q \ll 1$ is consistent with our asymptotic analysis in Sec.~\ref{sec:travelling}, where we find that the magnetization vanishes at leading order. Inserting \eqref{A3_isolated} into \eqref{equ:sin_pert} and rearranging the resulting $2\times 2$ system in terms of $A_a \equiv A_2$ and $A_0 = A_1 - A_2$ we find that $\tilde \lambda$ for $q\ll 1$ solves
\begin{equation}\label{equ:W_matrix_reduced}
 \widetilde W  \begin{pmatrix}
 	A_a\\ A_0
 \end{pmatrix} = \tilde \lambda \begin{pmatrix}
 	A_a\\ A_0
 \end{pmatrix},
\end{equation}
where $\widetilde W = - q^2 M(H + \frac{\pe^2}{2} G_\text{NR})$ and $M, H , G_\text{NR}$ coincide with the matrices in (\ref{eq:MH_out},\ref{equ:mf-M-full}) corresponding to the nonreciprocal form of the outer problem. Hence, the matrix of linear stability in the limit $q\ll 1$ coincides, up to the factor $-q^2$, with the matrix product $M G$ when writing the outer problem as a cross-diffusion problem (Example \ref{example3}).

\begin{figure}[thb]
\centering
\includegraphics[width=.5\linewidth]{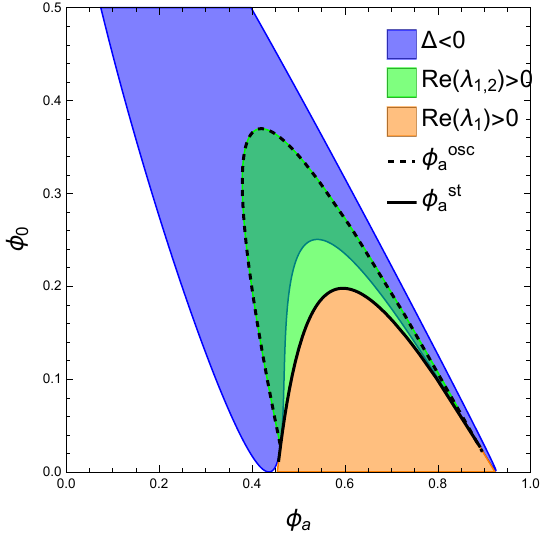}
\caption{\textbf{Classification of stationary and oscillatory instabilities. }
Stationary and oscillatory-type instabilities in terms of $\phi_a$ and $\phi_0$ for $\pe = 7.5$. Comparison between the prediction from the $3\times 3$ matrix $W$, namely $\min_q \phi_a^\text{osc}$ and $\min_q\phi_a^\text{st}$ using \eqref{equ:lin_stab_re} and \eqref{equ:lin_stab_comp} (black solid and dashed lines) and the $2\times 2$ matrix $\widetilde W$ corresponding to the $q\to 0$ limit (coloured regions). 
Denoting by $\lambda_1, \lambda_2$ the eigenvalues of $\widetilde W$, with Re$(\lambda_1) \ge$ Re$(\lambda_2)$, the blue region indicates complex eigenvalues (with determinant $\Delta = (\lambda_1+\lambda_2)^2 - 4 \lambda_1 \lambda_2 <0$), the organge region corresponds to one eigenvalue with positive real part (Re($\lambda_1) > 0 >$ Re($\lambda_2$)) and the green region to both eigenvalues with positive real part (Re$(\lambda_{1,2})>0$). Where the green region overlaps with the blue region, the instability is of oscillatory type.}
\label{fig:comparison}
\end{figure}

Fig.~\ref{fig:comparison} overlays the spinodal calculated according to \eqref{equ:lin_stab_re} and \eqref{equ:lin_stab_comp} with the approximation for small $q$ or outer region solution for $\pe = 7.5$. We find excellent agreement between the approximation and the full calculation, indicating that for this value of $\pe$, our model always goes unstable through a large-scale instability (either stationary or oscillatory). This is because, for $\pe \lessapprox 8$, $\inf_q \phi_a^{{\rm osc}}(q) = \phi_a^{{\rm osc}}(0)$.

\section{Numerical methods for the hydrodynamic PDEs}\label{app:numerical_methods}

\subsection{Time-dependent solutions} \label{sec:timedep}

We use a first-order finite-volume scheme to obtain one-dimensional numerical solutions $\rho_\sigma(x,t)$ to \eqref{equ:main} for $\sigma \in \{+1, -1, 0\}$. We first rewrite the equation in the form
\newcommand{\divz}{{\nabla_\zeta \cdot}}
\begin{align} \label{equ_almost_gradient}
		\pt \rho_\sigma + \p_x \left(M_{\sigma} \p_x U_{\sigma}  \right) + \sigma m = 0,
\end{align}
where $M_{\sigma}$ are scalar mobilities and $U_\sigma$ are scalar velocities. The mobilities are defined by $M_\sigma=\ds(\rho)\rho_\sigma $, and the velocities are given by
\begin{equation}
		 U_\sigma = - \Big[ \partial_x \log \rho_\sigma   + 
		 \partial_x Q( \rho ) \Big] + \pe 
		 \Big[ \sigma 
		 + \frac{ \pol s(\rho )}{ \ds(\rho) } 
		 \Big],
\end{equation}
where $Q: [0,1] \to \RR$ is such that $Q'(x) = \DD (x)/ \ds(x)$. \eqref{equ_almost_gradient} is complemented with periodic boundary conditions on $[0,L]$. 
We note that \eqref{equ_almost_gradient} is not a Wasserstein gradient flow as the velocities  $U_\sigma$ cannot be written as derivatives of an entropy (unless $\pe = 0$).

We discretize the spatial domain $[0,L]$ into $N$ cells of length $\Delta x = L/N$ and centre $x_i = (i+1/2)\Delta {x}$ for $i = 0, \dots, N-1$. 
We then approximate $\rho_\sigma(x_i, t)$ by the cell averages
\begin{equation}
	\rho_{\sigma, i}(t) = \frac{1}{ \Delta x} \int_{C_{i}} \rho_\sigma(x,t) \dd x. 
\end{equation}
We use the finite-volume scheme 
\begin{align} \label{equ_ODE_system}
\frac{\dd}{\dd t} \rho_{\sigma, i} = 
		&- \frac{F_{\sigma,i+1/2}-F_{\sigma,i-1/2} }{\Delta x} - \sigma m_i.
\end{align}
for $ i = 0, \dots, N-1$, with $F_{\sigma,-1/2} \equiv F_{\sigma,N-1/2}$ using periodicity. We approximate the flux $F_\sigma$ at the cell interfaces by the numerical upwind flux
\begin{equation} \label{num_upwind1}
	F_{\sigma,i+1/2} = d_s(\rho_{\sigma, i})\rho_{\sigma, i}(U_{\sigma,i+1/2})^+  + d_s(\rho_{\sigma,i+1})\rho_{\sigma,i+1} (U_{\sigma,i+1/2})^- ,
\end{equation} 
where $(\cdot)^+ = \max(\cdot, 0)$ and $(\cdot)^- = \min(\cdot, 0)$ and $\rho_{\sigma, N} \equiv \rho_{\sigma,0}$. The velocities $U_\sigma$ are approximated by centered differences
\begin{align}
			U_{\sigma, i+1/2} =&
	- \left[ 
	 \frac{\log \rho_{\sigma, i+1} - \log \rho_{\sigma, i} }{\Delta {x}}
	+\frac{Q( \rho_{i+1,j} ) - Q( \rho_{\sigma, i} ) }{\Delta {x}}
	\right]
+ \pe \left[ \sigma+
	\frac{1}{2} \left (   \frac{ \pol_{i+1}  s( \rho_{i+1})  }{  \ds (\rho_{i+1}) } + \frac{ \pol_{i} s( \rho_{i})  }{  \ds (\rho_{i}) } \right)   
	\right].
\end{align}
Finally, the resulting system of ODEs \eqref{equ_ODE_system} for $\rho_{\sigma,i}(t)$ is solved by the forward Euler method with an adaptive time stepping condition satisfying 
\begin{equation}\label{equ_cfl}
	\Delta t = \min \left \{ 10^{-5}, \Delta x/(6a) \right \}
\end{equation}
with $a = \max_{\sigma, i} \{ \vert U_{\sigma, i} \vert \}$.
In \cite{krukFiniteVolumeMethod2021}, a CFL condition of the form \eqref{equ_cfl} is shown to result in a positivity-preserving numerical scheme. In contrast to our model, their scheme is second-order in space, using a linear density reconstruction at the interfaces that preserve positivity. Here, we follow instead \cite{brunaPhaseSeparationSystems2022,masonExactHydrodynamicsOnset2023a} and use the values at the center of the cells. In our numerical tests, we observe \eqref{equ_cfl} to be sufficient to preserve positivity.

We initiate the scheme with a perturbation around the homogeneous state, $\rho_\sigma(x,0) = \phi_\sigma + \delta \tilde \rho_\sigma(x)$ with $\phi_\pm = \phi_a/2$ and $\phi_0 = \phi_p$. We normalise the perturbation so that $\Vert  (\tilde \rho_+,\tilde \rho_0, \tilde \rho_-) \Vert_2 =1$, where $\Vert u \Vert_2 = \left(\int_0^L |u |^2 \dd x \right)^{1/2}$ is the $L_2$ norm. For a uniformly random perturbation, we define 
\begin{equation}\label{rand_ic}
	\tilde \rho_\sigma^\text{rand} (x) \sim \text{Unif}[-1,1]. 
\end{equation}
We also use the eigenfunctions from linear stability analysis to generate perturbations. In particular, we solve \eqref{equ:sin_pert} for $q = 2 \pi /L$ and select the solution corresponding to the eigenvalue, $\lambda$, with the largest real part and nonnegative imaginary part.  
We define left and right traveling  perturbations,
\begin{equation}\label{left_right_ic}
	\tilde \rho_\sigma^\text{L} (x) = \text{Re}[ A_\sigma \exp(iqx) ], \quad \tilde \rho_\sigma^\text{R} (x) = \text{Re}[ A_\sigma \exp(-iqx) ], 
\end{equation}
 where $A_\pm = (A_2 \pm A_3)/2$, $A_0 = A_1-A_2$. (Both perturbations will be stationary when $\text{Im} \lambda =0$.) 
 In Figs. \ref{fig:MIPS}, \ref{fig:CP} we set $\tilde \rho_\sigma \propto \tilde \rho_\sigma^\text{rand}$ and $\delta = 0.1$ to mimic the random initial condition of the particle simulation. 
 In Fig. \ref{fig:TP} we set 
 $\tilde \rho_\sigma \propto \tilde \rho_\sigma^\text{L}
 +\tilde \rho_\sigma^\text{R}
 +\tilde \rho_\sigma^\text{rand}$ and $\delta = 0.1$. The left and right traveling perturbations, $\tilde \rho_\sigma^\text{L}
 +\tilde \rho_\sigma^\text{R}$, seeds the growth of counterpropagating interfaces, but the random perturbation, $\rho_\sigma^\text{rand}$, allows asymmetry to grow. Eventually, the solution reaches a steady, left-traveling TP state.
  In Fig. \ref{fig:steady_states_finite} (a) we set $\tilde \rho_\sigma \propto \tilde \rho_\sigma^\text{L}
 +\tilde \rho_\sigma^\text{R}$ and $\delta = 0.1$. This initial condition ensures that any left and right traveling interfaces will be balanced; therefore, the steady cannot be TP. On the other hand, in Fig. \ref{fig:steady_states_finite} (b) we set $\tilde \rho_\sigma \propto \tilde \rho_\sigma^\text{L}$ and $\delta = 0.1$. When $\lambda$ is complex, this encourages left-traveling TP steady states. 

\subsection{Traveling solutions in the finite system} \label{sec:travelling_num}

\newcommand{\uu}[0]{\boldsymbol{\varrho}}

We seek one-dimensional traveling solutions $\varrho_\sigma(z)$ to \eqref{equ:travelling} with periodic boundary conditions, where $z = x-ct/L \in [-L/2,L/2]$. Integrating \eqref{equ:travelling} gives
\begin{align} \label{newsystem_travel}
	-(c/L) \varrho_\sigma  + M_\sigma \partial_z U_\sigma + \sigma \int_{-L/2}^z \poltravel(y) \dd y =  A_\sigma,
\end{align}
where $A_\sigma$ are integration constants and the mobilities $M_\sigma$ and velocities $U_\sigma$ are as given in Subsec.~\ref{sec:timedep} but replacing $\rho_\sigma$ by their traveling frame counterparts $\varrho_\sigma$. 
Additionally, we have the mass constraints
\begin{equation}
	\frac{1}{L}\int_{-L/2}^{L/2} \varrho_\sigma \dd z = \phi_\sigma,
\end{equation}
with $\phi_\pm = \phi_a/2$ and $\phi_0 = \phi_p$. [Steady-state solutions must have $\int \poltravel \dd z = 0$, as seen by integrating \eqref{equ:travelling}.]

We use the same finite-volume spatial discretization as in Subsec.~\ref{sec:timedep}, resulting in the discretized equations
\begin{align} \label{discrete_travel}
		-\frac{c}{2L} \left (\varrho_{\sigma, i} + \varrho_{\sigma, i+1} \right) +  F_{\sigma,i+1/2}  + \sigma \Delta z \sum_{j = 0}^{i} \poltravel_j &= A_\sigma,	
\end{align}
for $i =0, \dots N-1$, where $F_{\sigma, i+1/2}$ is defined in \eqref{num_upwind1} (but in terms of the travelling frame variables), together with mass constraints
\begin{align} \label{mass_constraints}
	\Delta z \sum_{i = 0}^{N-1} \varrho_{\sigma, i} = L\phi_\sigma, 
\end{align}
So far, we have $3N$ equations ($3(N-1)$ in \eqref{discrete_travel} and three in \eqref{mass_constraints}) for $3N + 1$ unknowns. The remaining degree of freedom is removed by noting the problem has translational symmetry; hence, we set $\varrho_{0,0} = \phi_p$ without loss of generality.

The nonlinear system of $3N$ equations (\ref{discrete_travel})-(\ref{mass_constraints}) is then solved numerically using \texttt{NonlinearSolve()} from the \texttt{Julia} \texttt{NonlinearSolve.jl} package  \cite{pal2024nonlinearsolve}, with parameters \texttt{reltol=1e-8}, \texttt{abstol=1e-8} and \texttt{maxiters=20}. The solver first tries less robust Quasi-Newton methods for more performance and then tries more robust techniques if the faster ones fail. As such, it requires a close enough initial guess for convergence. The first time, we initialize the iterative solver with the long-time ($T=1000$) solution of the time-dependent problem (Subsec.~\ref{sec:timedep}) with parameters $\phi=0.67, \phi_a=0.37$,$\pe = 7.5$, $L=25$,$N=500$. Once we have found a traveling solution to (\ref{discrete_travel},\ref{mass_constraints}), we use it as an initial condition for the problem with slightly altered parameters: $\phi_a' = \phi_a\pm0.01$, $\phi_p' = \phi_p\pm0.01$, as well as larger domains, e.g., $L'=2 L$ or $N'= 2 N$.

\subsection{Traveling solutions in large systems}
We seek solutions $\varrho_\sigma(z)$ to \eqref{equ:outer_leading} for $z\in [-L/2, L/2]$ with periodic boundary conditions with up to two interfaces (inner regions) at $z=0$ and $z = \delta<L/2$. In what follows, it is convenient to consider the domain $z\in [0, L]$ instead so that one of the interfaces (if any) is at the interval ends.

We define three outer problems depending on the number of interfaces. Each of them takes different input parameters:
\begin{itemize}
    \item The no interface problem (\texttt{outer0}) takes the total volume fractions $\phi, \phi_a$.
    \item The one interface problem (\texttt{outer1}) takes the total volume fraction $\phi$ and the tie-line parameter $\nu$.
    \item The two interfaces problem (\texttt{outer2}) takes one tie-line parameter $\nu$ and the separation between interfaces $\delta$.
\end{itemize}

To solve the \texttt{outer0} problem, we write \eqref{equ:outer_leading} in terms of $\varrho$ and $\varrho_a$ and integrate, leading to
\begin{align} \label{travelling_outer}
\begin{aligned}
	-(c/L) \varrho + F &= A_1, & \qquad F & = - \varrho' - \tfrac{\pe^2}{2}(1-\varrho)\partial_z[d_s(\varrho) \varrho_a]\\
	-(c/L) \varrho_a + F_a & = A_2, & F_a &= - d_s(\varrho) \varrho_a' - \varrho_a \DD(\varrho) \varrho' - \tfrac{\pe^2}{2}\left [ \varrho_a s(\varrho) + d_s(\varrho) \right] \partial_z [d_s(\varrho) \varrho_a],
\end{aligned}	
\end{align}
where $A_1,A_2$ are constants of integration (note that we only have two equations since $\poltravel =0$ at leading order). Additionally, we have the mass constraints 
\begin{equation} \label{massouter}
	\frac{1}{L} \int_0^L \varrho \dd z = \phi, \qquad \frac{1}{L}\int_0^L \varrho_a \dd z = \phi_a,
\end{equation}
We discretize the domain $[0, L]$ into $N$ intervals of equal length $\Delta z = L/N$, and solve for $\varrho_i$ and $\varrho_{a,i}$ at grid points $z_i =   i \Delta z$ for $i = 0, \dots, N-1$. 
The discretized equations of \eqref{travelling_outer} are, for $i = 0, \dots, N-1$,
\begin{align} \label{discrete_travel2}
\begin{aligned}
		-\frac {c}{2L} ( \varrho_{i}+ \varrho_{i+1}) +  F_{i+1/2}  = A_1,\qquad -\frac {c}{2L}( \varrho_{a,i} +\varrho_{a,i+1}) +  F_{a,i+1/2} = A_2,
	\end{aligned}
\end{align}
with the fluxes at the half-points approximated by centered differences, e.g.,
$$
F_{i+1/2} = - \frac{1}{\Delta z}(\varrho_{i+1}-\varrho_i) - \frac{\pe^2}{4 \Delta z} (2 - \varrho_i - \varrho_{i+1}) \left[d_s(\varrho_{i+1}) \varrho_{a,i+1} - d_s(\varrho_{i}) \varrho_{a,i} \right],
$$
with $\varrho_N = \varrho_0$ and $\varrho_{a,N} = \varrho_{a,0}$ using periodicity. The mass constraints \eqref{massouter} become
\begin{align} \label{mass_constraints2}
	\Delta z \sum_{i = 0}^{N-1} \varrho_i = L\phi, \qquad  \Delta z \sum_{i =0}^{N-1} \varrho_{a,i} = L\phi_a. 
\end{align}
This yields a set of $2N + 3$ unknowns ($\varrho_i, \varrho_{a,i}, c, A_1, A_2$) and $2N + 2$ equations [Eqs. (\ref{discrete_travel2}) and (\ref{mass_constraints2})]. As in Subsec.~\ref{sec:travelling_num}, the additional constraint comes from noting that the problem has translational symmetry; we fix $\varrho_0 = \phi$.

\begin{figure*}
\centering
\includegraphics[width=17.8cm]{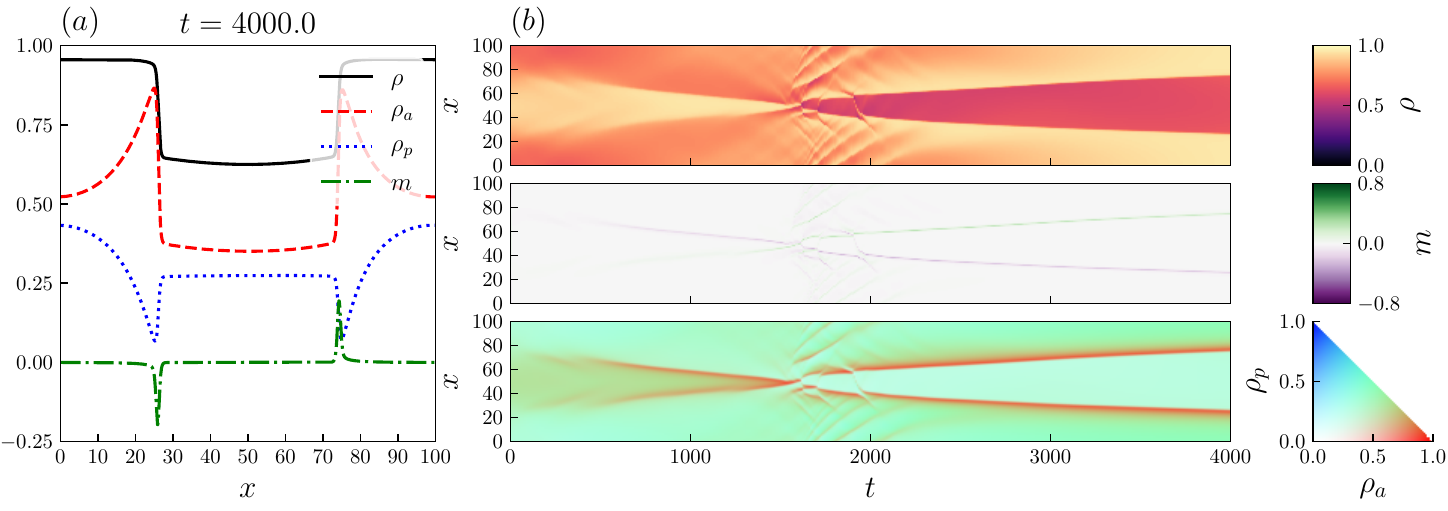}
\caption{\textbf{Scattering in a counter-propagating solution. }
CP solution to \eqref{equ:main}. The initial condition is the homogeneous state ($\rho_\pm = \phi_a/2, \rho_0 = \phi_p$) perturbed by a uniform random perturbation (see Sec.~\ref{sec:timedep} for details).
(a) Density profile at $t=4000$. 
(b) Kymographs showing the spatiotemporal dynamics. 
Parameters: $\pe = 7.5$, $L =100.0$, $\phi_a = 0.5$, $\phi_p = 0.3$, $\Delta x = 0.05$.
}
\label{fig:CP_large}
\end{figure*}

\begin{figure*}
\centering
\includegraphics[width=11cm]{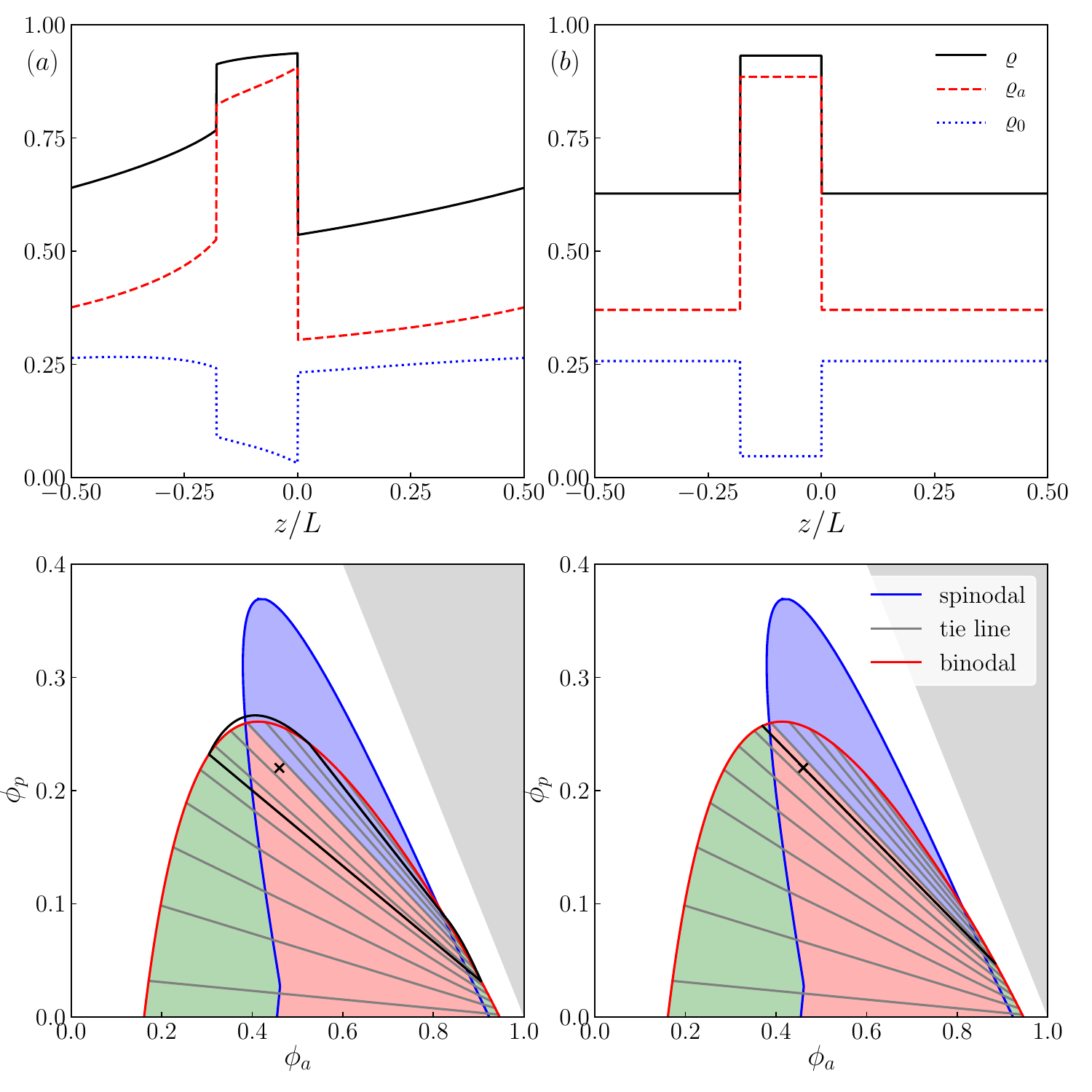}
\caption{\textbf{Multi-stability of solutions.}
Two solutions, both obtained from Eq.~(8) with the same parameters.   (a) T solution.  (b) PS solution.  
The top panels display the solution profile as a function of $z/L$. The bottom panels display the profile (solid black) overlayed on phase diagrams spanned by $\phi_a$ and $\phi_p$. 
 The Spinodal (Blue) encloses the region of linear stability of homogeneous solutions, and the Binodal (Red) encloses the region of Phase separation. 
 Black crosses mark the volume fraction.
Parameters: $\pe = 7.5$, $\phi_a = 0.4608$, $\phi_p = 0.2201$, $N = 1024$}
\label{fig:multistability}
\end{figure*}

\begin{figure*}
\includegraphics[width=11cm]{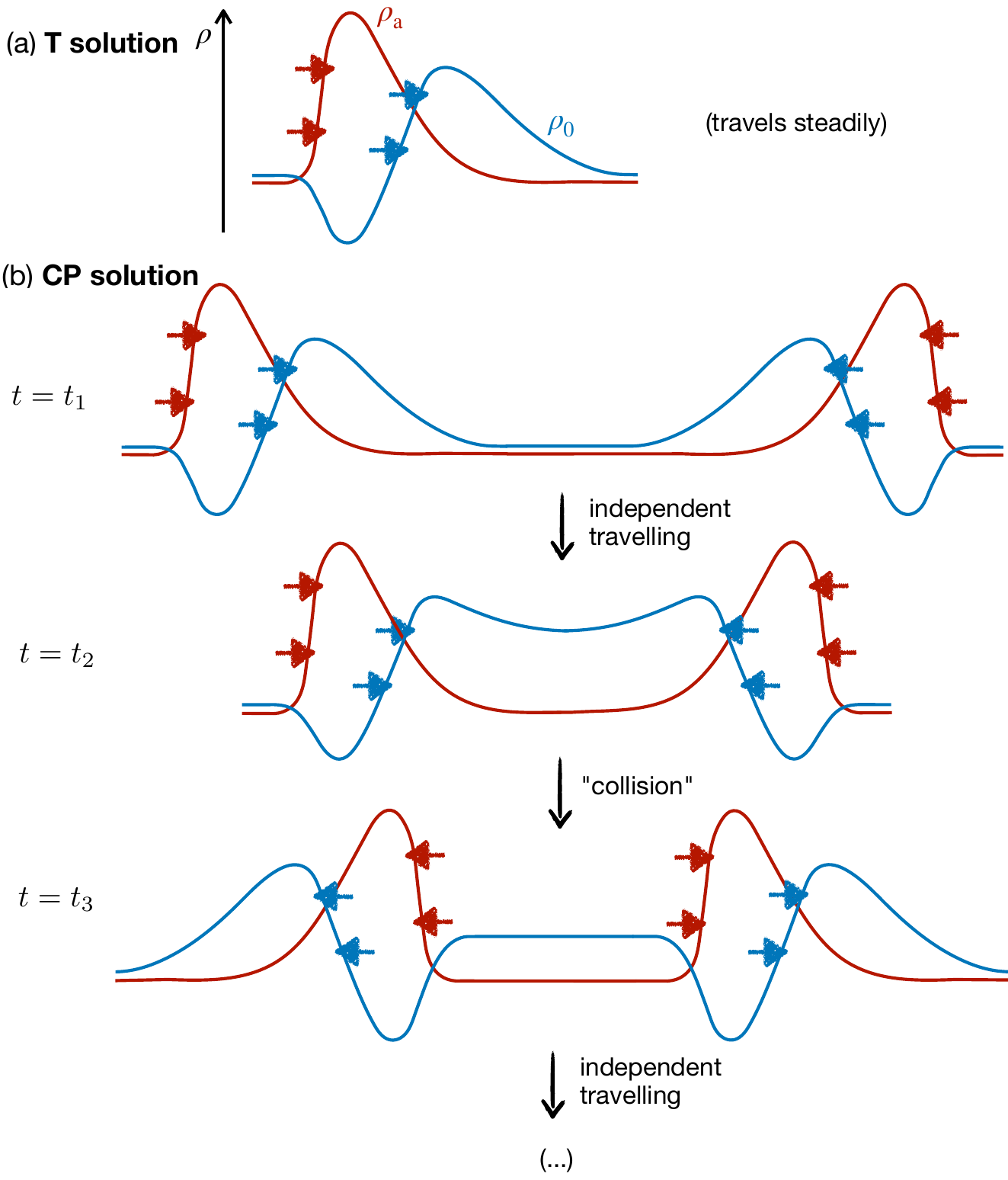}
\caption{\textbf{Illustration of T and CP solutions. }
(a) Sketch of T solution with positive velocity, showing density of active and passive particles as a function of position (see Fig.~\ref{fig:TP} of main text for a similar solution with negative velocity). The red arrows indicate the self-propulsive force acting on the active particles.  The blue arrows indicate forces applied by the active particles on the passive ones (via the cross-diffusion terms in the equation of motion).  This generates the positive propagation.  (b) Sketch of a CP solution at three different times in a large system (see Fig.~\ref{fig:CP} of main text for a similar solution in a smaller system).  Initially the CP solution is well-approximated by a combination of two T solutions with opposite velocities of equal magnitude.  The two propagating clusters approach each other and eventually collide.  After the collision process one again sees a combination of two T solutions which travel away from each other.  The cycle will eventually repeat due to periodic boundaries.}
\label{fig:sketch-tcp}
\end{figure*}

For the \texttt{outer1} problem with parameters $\phi$ and $\nu$, we first solve the coexisting phases problem (Sec.~\ref{sec:coexist}) given $\nu$. This results in vapor and liquid phases $\phi_v(\nu)$ and $\phi_l(\nu)$, respectively, as well as individual components $(\varrho_\sigma)_v$ and $(\varrho_\sigma)_l$ in each phase. They satisfy $(\varrho_\sigma)_v 
\equiv (\varrho_+, \varrho_-, \varrho_0)_v = (\tfrac12\phi_{a,g}, \tfrac12 \phi_{a,g}, \phi_{p,g})$ and similarly for $(\varrho_\sigma)_l$.
	If $\phi_v(\nu) = \phi_l(\nu)$, it means there is no inner region and \texttt{outer1} has no solution (and we revert to \texttt{outer0}). 
Next, we solve the outer problem \eqref{travelling_outer} with Dirichlet boundary conditions $\varrho(0) = \phi_v, \varrho(L) = \phi_l, \varrho_a(0) = \phi_{a,g}$ and $\varrho_a(L) = \phi_{a,l}$. 
In contrast to \texttt{outer0}, we now have $2(N-1) + 3$ unknowns ($\varrho_i, \varrho_{a,i}$ for $i = 1, \dots, N-1$ and $c, A_1, A_2$) and $2N + 1$ equations [Eqs. (\ref{discrete_travel2}) for $i = 0, \dots, N-1$ and one mass constraint (\ref{mass_constraints2})]. The actives volume fraction $\phi_a$ is found a posteriori as $\phi_a = (1/N) \sum_{i =0}^{N-1} \varrho_{a,i}$. 

To obtain a solution to the \texttt{outer2} problem with two interfaces at a distance $\delta$ and the first one with a tie-line parameter $\nu$, the first step is as in \texttt{outer1}: we solve the coexisting phases problem with $\nu$ and this gives us Dirichlet boundary conditions at $z = 0$ and $z = L$ (as before, if $\phi_v(\nu) = \phi_l(\nu)$ there is no interface and we revert to \texttt{outer1}). We take the separation between interfaces $\delta$ to be $\delta = (j + 1/2) \Delta z $ for some $j$, that is, the second interface lies at the half-grid point $z_{j+1/2}$. To force a second interface between the grid points $z_j$ and $z_{j+1}$, the two equations (\ref{discrete_travel2}) corresponding to this half-grid point are replaced by three binodal equations (Sec.~\ref{sec:coexist})
\begin{equation}\label{second_interface}
\Phi(R_{j})  - R_{j} \Phi'(R_{j}) = \Phi(R_{j+1})  - R_{j+1} \Phi'(R_{j+1}),\qquad \Phi'(R_{j}) = \Phi'(R_{j+1}), \qquad \nu_j = \nu_{j+1},
\end{equation}
where we introduced the shorthand $\nu_j = \frac{\varrho_j-\varrho_{a,j}}{1-\varrho_j}$ and $R_j = R(\varrho_j, \nu_j)$. 
Next, we solve the modified outer problem Eqs. (\ref{discrete_travel2})-(\ref{second_interface}) with the Dirichlet boundary conditions coming from the first interface and one additional constraint coming from the second interface -- this explains why, in this case, the total volume fraction $\phi$ is not an input parameter but obtained a posteriori together with $\phi_a$.

The nonlinear systems \texttt{outer0},\texttt{outer1},\texttt{outer2} are again solved numerically using \texttt{NonlinearSolve()} with paramters \texttt{reltol=1e-8}, \texttt{abstol=1e-8} and \texttt{maxiters=20}.
We first initialise \texttt{outer0} using the finite domain solution (Subsec.~\ref{sec:travelling_num}) with $\phi=0.80, \phi_a=0.45$, $\pe = 7.5$, $L=100$, $N=3200$. Once an initial solution to \texttt{outer0} has been found, we initialize the iterative solver using an existing solution with similar parameters:  $\phi_a' = \phi_a\pm0.01$, $\phi_p' = \phi_p\pm0.01$. 
We first initialise \texttt{outer1} using the finite domain solution (Subsec.~\ref{sec:travelling_num}) with $\phi=0.8, \phi_a=0.3$,$\pe = 7.5$,$N=3200$. Once an initial solution to \texttt{outer0} has been found, we initialize the iterative solver using an existing solution with similar parameters:  $\phi' = \phi\pm0.01$, $\nu' = \nu \pm0.01$.
We first initialise \texttt{outer2} using the 
\texttt{outer1} solution with $\phi=0.65, \nu=0.36$,$\pe = 7.5$,$N=1024$, and place the new interface at $j=299$. Once an initial solution to \texttt{outer2} has been found, we initialize the iterative solver using an existing solution with similar parameters:  $\nu' = \nu \pm0.01$, $j' = j\pm8$.

\section{CP and T solutions} \label{sec:diagrams}

Fig.~\ref{fig:CP_large} illustrates a CP solution in a large domain, showing the emergence of narrow interfaces as in T states.  Fig.~\ref{fig:multistability} shows an example state point where both T and PS solutions can exist -- this is multistability, as discussed in Sec.\ref{sec:multistab} of main text.

To further clarify the relationship between CP and T solutions, we sketch in Fig.~\ref{fig:sketch-tcp} the qualitative behavior that we expect in large systems.  The T solution involves a localized set of active particles that ``push'' a set of passive ones through the system (Fig.~\ref{fig:sketch-tcp}(a)).  Red arrows indicate the positive self-propulsive forces that act in the (inner) region where $\rho_a$ rises steeply; $m$ is positive here and the relevant terms in the equation of motion are those proportional to Pe.  Blue arrows indicate that the passive particles are pushed by the active ones, primarily due to cross-diffusion terms in the equation of motion.

The CP solution consists of two localized sets of active particles moving in opposite directions, resembling a combination of two T solutions traveling in opposite directions (Fig.~\ref{fig:sketch-tcp}(b)). There are two associated inner regions where these particles propel themselves (red arrows), and the active particles push the passive ones as before. When the two traveling objects approach each other, they collide. The collision process is complex, with the clusters departing the collision at high speeds. (Note: we depict the traveling objects as localized for illustrative purposes, but the actual densities are functions of $x/L$ in the outer region.) 

An essential feature of a stable CP solution is that both traveling objects survive the collision. In contrast, while the early-time behavior shown in Fig.~\ref{fig:TP} resembles a CP solution, after three collisions, it becomes evident that the two traveling objects have unequal masses, resulting in only one object surviving after the fourth collision. The outcome for large times corresponds to a T solution.

\bibliography{article_refs.bib}

\end{document}